\documentclass[fleqn,10pt]{wlscirep}
\usepackage[utf8]{inputenc}
\usepackage[T1]{fontenc}
\usepackage{bm}
\usepackage{soul}

\usepackage{graphicx,xcolor}
\usepackage{subcaption}

\newcommand{\TO}[0]{TiO$_2$}
\newcommand{\STO}[0]{SrTiO$_3$}

\newcommand{\tisa}[0]{Ti$_{S0}^A$}
\newcommand{\tisb}[0]{Ti$_{S0}^B$}
\newcommand{\tisia}[0]{Ti$_{S1}^A$}
\newcommand{\tisib}[0]{Ti$_{S1}^B$}

\newcommand{\tisiib}[0]{Ti$_{S2}^B$}
\newcommand{\EPOL}[0]{$E_{\rm pol}$}
\newcommand{\barEPOL}[0]{$\Bar{E}_{\rm pol}$}
\newcommand{\EPOLML}[0]{$E_{\rm pol}^{\rm ML}$}
\newcommand{\EPOLDFT}[0]{$E_{\rm pol}^{\rm DFT}$}
\newcommand{\barEPOLDFT}[0]{$\Bar{E}_{\rm pol}^{\rm DFT}$}
\newcommand{\barEPOLML}[0]{$\Bar{E}_{\rm pol}^{\rm ML}$}
\newcommand{\ie}[0]{\textit{i.e.},}
\newcommand{\eg}[0]{\textit{e.g.},}

\newcommand{\vo}[0]{$V_{\rm O}$}
\newcommand{\VO}[0]{$V_{\rm O}$}
\newcommand{\cvo}[0]{$c_{V_{\rm O}}$}
\newcommand{\cnb}[0]{$c_{\rm Nb}$}

\title{Machine Learning for Exploring Small Polaron Configurational Space}

\author[1,2]{Viktor C. Birschitzky}
\author[1]{Florian Ellinger}
\author[2]{Ulrike Diebold}
\author[1]{Michele Reticcioli}
\author[1,3]{Cesare Franchini}
\affil[1]{University of Vienna, Faculty of Physics and Center for Computational Materials Science, Vienna, Austria}
\affil[2]{Institute of Applied Physics, Technische Universität Wien, 1040 Vienna, Austria}
\affil[3]{Department of Physics and Astronomy 'Augusto Righi', Alma Mater Studiorum - Universit\`{a} di Bologna, Bologna, 40127 Italy}

\affil[*]{e-mail: cesare.franchini@univie.ac.at}

\begin{abstract}
Polaron defects are ubiquitous in materials and play an important role in many processes involving carrier mobility, charge transfer and surface reactivity.
Determining the spatial distribution of small polarons is essential to understand materials properties and functionalities.
This requires an exploration of the configurational space, which is computationally demanding when using standard first principles methods, and technically prohibitive for many-polaron systems.
Here, we propose a machine-learning (ML) accelerated search that compares the energy stability of different polaron patterns and determines the ground state configuration.
The kernel-regression based ML model is trained on databases generated by density functional theory (DFT) calculations on a minimal set of initial polaron patterns, obtained by using either molecular dynamics simulations or a random sampling approach.
To establish an efficient mapping between training data and configuration stability we designed simple descriptors that model the interactions among polarons and charged point defects.
The proposed DFT+ML protocol is used here to explore millions of polaron configurations for two different systems, oxygen defective rutile \TO(110) and Nb-doped \STO(001).
Our data shows that the ML-aided search correctly individuates the ground-state polaron patterns, proposes polaronic configurations not visited in the training and can be used to efficiently determine the optimal distribution of polarons at any charge concentration.
\end{abstract}
\begin{document}

\flushbottom
\maketitle

\thispagestyle{empty}

\section{Introduction}
Polarons are quasiparticles that can form in polarizable materials by entanglement between charge carriers and lattice distortions~\cite{Franchini2021, Alexandrov2010book}.
An unbound electron or hole injected in a material can interact with the lattice vibrations (phonon).
If the electron-phonon coupling is strong enough, the excess carrier induces local structural deformations associated with a potential well in which the carrier is selftrapped~\cite{LANDAU1933, Zienau2}.
Polarons represent a long-standing yet still exciting field of research with profound impact in different disciplines ranging from physics to chemistry and material science, where they play a pivotal role in several phenomena of practical importance such as carrier mobility~\cite{Coropceanu2007, PhysRevLett.110.196403, PhysRevLett.91.256403, Deskins2009a, PhysRevLett.122.096101, PhysRevLett.114.146401,PhysRevLett.123.076601}, surface reactivity~\cite{Papageorgiou2391, PhysRevLett.122.016805, PhysRevX.7.031053, Rousseau2020, Sokolovic14827} and optical excitations~\cite{Mechelen2008STO, Yoon1998, Klimin_2020, doi:10.1021/acs.jpclett.9b02342}, and represent a testbed for the development of many theoretical models and numerical methods~\cite{Franchini2021}.
    
Small polarons, whose wave function is spatially confined within a few~\AA\ around their trapping site, are subjected to thermally-activated hopping processes, which enable polaron mobility.
As a consequence, small polarons can travel through the material forming different spatial distributions (polaron configurations) that have a strong impact on the properties and functionality of the material~\cite{PhysRevB.98.045306}.
Predicting favorable polaron configurations is key to correctly interpret experimental measurements and predict material's behaviour, but it is a challenging task.
Polaron ground-state distributions result from the balance between contrasting interactions, primarily polaron-polaron repulsion and the attraction between the negatively-charged polarons and the positively-charged donor defects (\eg\ oxygen vacancies and/or dopants)~\cite{Reticcioli2019}.
Considering that polaron formation is favored on surface or near-surface sites, dimensionality effects and surface reconstruction also play a crucial role in defining the optimal polaron configuration, complicating the scenario even further~\cite{PhysRevX.7.031053, PhysRevB.98.045306}.

Material-specific properties of small polarons are typically computed by using first principles approaches in the framework of density functional theory (DFT) and appropriate extensions~\cite{Shluger_1993, Reticcioli2019, PhysRevLett.122.246403, PhysRevB.99.235139}. 
The DFT modelling of defects-induced polarons is complicated by the need to adopt large supercells in order to attenuate artificial interactions between periodic images of the polaron, which hampers an efficient exploration of the huge configurational space and makes the calculations computationally very demanding~\cite{GOYAL20171,PhysRevB.51.4014,PhysRevLett.102.016402}.
    
The routine approach to explore the polaron configurational space is based on a combination of manual selection and molecular dynamics (MD)~\cite{Reticcioli2019,PhysRevLett.105.146405, occupation_matrix_control}.
This protocol involves a three-step procedure:
(\textit{i}) Generating a pool of initial structures, where the polaron trapping sites are manually selected (using different types of site-controlled strategies~\cite{Reticcioli2019, occupation_matrix_control, doi:10.1021/acs.jctc.0c00374});
(\textit{ii}) subsequent MD runs at temperatures high enough to activate small polaron hopping, thus allowing for a guided exploration of energetically favorable configurations;
(\textit{iii}) finally, a set of ground-state static DFT calculations for all inequivalent configurations found in the MD runs will determine the energetically favorable solutions at low temperature.
This DFT-MD scheme can be easily automatized in a workflow, but the prohibitively long time required to execute MD runs on large supercells and the sporadic thermally-activated polaron hopping events, prevent an efficient exploration and restricts \emph{de facto} the search to a limited subset of the full configurational space.\\
An alternative approach to bypass the extensive MD-simulations 
(step~\textit{ii}) 
would be to create a larger pool of manually-selected polaronic configurations in step~\textit{i} and directly determine their energy stability as in step~\textit{iii}.
However, this strategy relies on human intuition that could bias the selection process of trapping sites, thus excluding possible favorable configurations from the investigated pool.
Exhaustive searches including all possible configurations can be performed for the simplest cases, \ie\ the dilute limit, where the amount of polarons is relatively low.
At higher defect concentrations, however, the combinatorial growth of configurations is excessive (\ie\ dense limit).
Therefore, random sampling approaches have to be used, which cannot guarantee the determination of the proper ground state.
For this reason, alternative sampling methods are necessary to correctly describe the properties of materials hosting small polarons.
    
In this work we propose a data-driven strategy for an accelerated exploration of the polaron configurational space in order to predict the optimal trapping-site patterns for polarons.
In recent years, machine learning (ML) has been used extensively in problems 
involving many combinatorial possibilities~\cite{schmidt_recent_2019}, as in the cases of chemical compound space~\cite{von_lilienfeld_retrospective_2020}, materials design \cite{ramprasad_machine_2017} and the calculation of potential energy surfaces and force fields~\cite{behler_representing_2007, jinnouchi_--fly_2019, bartok_gaussian_2010}.
Here, we propose a simple kernel regression scheme~\cite{bishop_pattern_2006,schutt_machine_2020}, with descriptors that embody the polaron-polaron and polaron-defects interactions, trained on minimal DFT datasets to assess the relative stability of the initial (few hundreds) polaron configurations. 
The trained ML algorithm can then be used to extend the exploration of the configurational space by systematically analyzing millions of configurations, going far beyond the limits of \emph{first principles} MD or any alternative sampling approaches based entirely on DFT.
    
The proposed DFT+ML strategy is applied here to two prototypical polaronic materials considering different types of doping: the oxygen-defective rutile TiO$_2$(110) surface~\cite{Rousseau2020, diebold_surface_2003, Reticcioli2019, PhysRevLett.113.086402, PhysRevLett.105.146405, PhysRevB.75.195212, Papageorgiou2391, Moses2016} and the Nb-doped perovskite SrTiO$_3$(001) surface~\cite{PhysRevB.95.035301, PhysRevB.90.085202, doi:10.1142/S0217979214300096}.
We show that the ML-aided search correctly identifies the ground-state polaron configuration for arbitrary carrier density (from the dilute to the dense limit), as confirmed by benchmark DFT tests on selected ML-predicted configurations.
Our results show that our proposed method can be employed to efficiently determine the optimal polaron patterns in diverse polaronic materials, and can be trained to account for the interaction of polarons with different point-like defects (oxygen vacancies and dopants, but also interstitials and adatoms for example).

    \begin{figure}[ht!]
        \centering
        \includegraphics[width=1.0\textwidth]{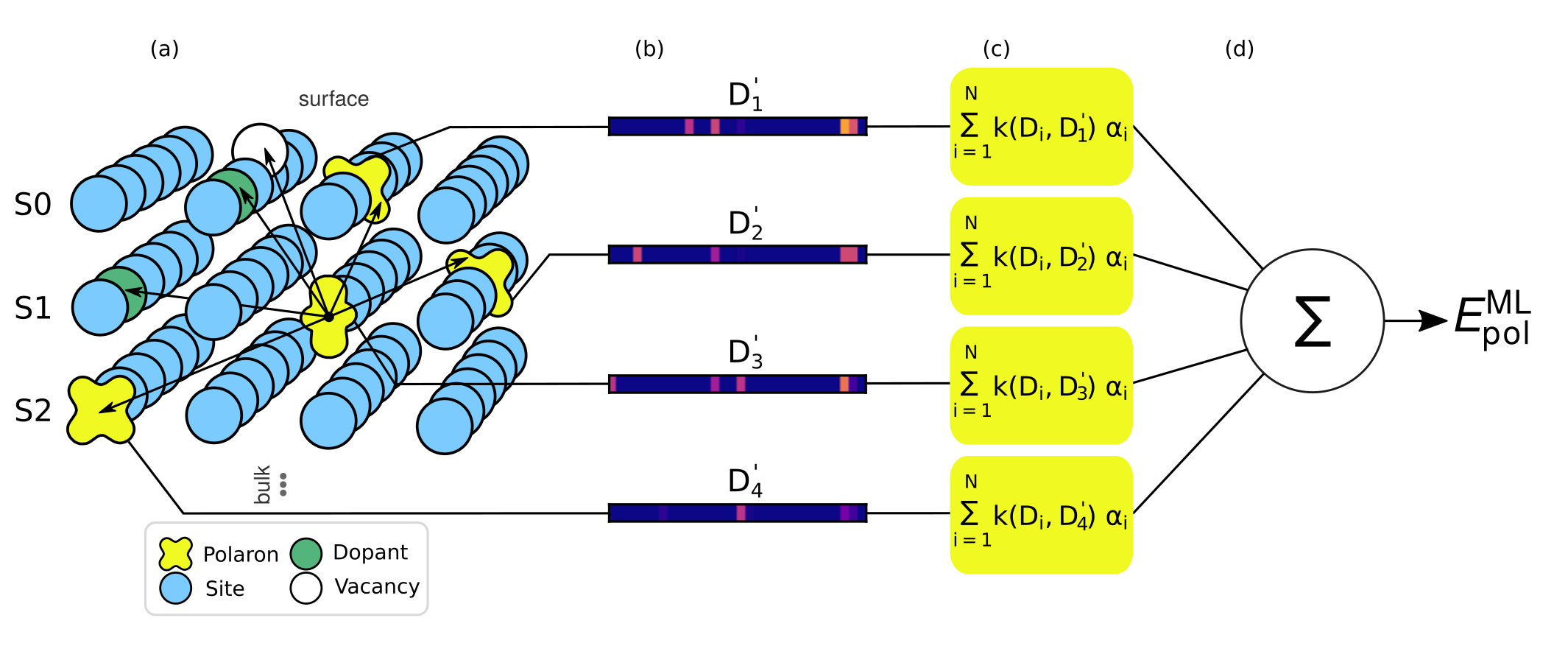}
        \caption{
        Schematic representation of the ML model. \textbf{(a)} Sketch of a polaron configuration in  a generic atomic structural model. Possible polaron-trapping sites (blue circles), defects (dopants and atom vacancies, green and white circles, respectively) and polaron charge densities (yellow) are shown. The local interactions of the the reference polaron with the defects and other polarons are indicated by arrows.
        \textbf{(b)} Descriptor representation $D'_i$ for every polaron in the configuration. The strength of the polaron-polaron and defect-polaron interactions are represented by colour gradient (from blue to pink, details in Sec.~\ref{method:desc} and Section S3 of the SM).
        \textbf{(c)} Kernel functions $k(\cdot,\cdot)$ with parameters $\alpha_i$ measure the similarity of the target-polaron descriptor $D'$ with the training-set samples $D_i$, and assign a virtual single-polaron energy to every polaron in the configuration.
        \textbf{(d)} The ML-predicted single-polaron energies are summed up to return the total polaronic energy \EPOLML\ of the configuration, which can be compared to the \EPOLDFT.
        }
        \label{fig:algo}
    \end{figure}

\section{Results}

We start with a brief description of the ML-aided algorithm designed to predict the stability of multipolaron configurations (more details can be found in the Methods~\ref{sec:method}).
The schematic protocol, shown in Fig.~\ref{fig:algo},
involves a mapping between a general surface structure containing charge-donor defects (e.g. dopants or oxygen vacancies, Fig.~\ref{fig:algo}(a)) and possible polaron hosting sites (typically transition metal ions) with a kernel-regression scheme. The connection is established by means of a simple descriptor ($D$) of local interactions (Fig.~\ref{fig:algo}(b), details provided in Section~\ref{method:desc}) which encode the spatial distance between the reference polaron to other polarons and point defects included within a cutoff sphere.
This representation is used in a kernel regression scheme (Fig.\ref{fig:algo}(c)) to map a given many-polaron configuration to the corresponding polaron formation energy (\EPOL) as calculated by DFT (see Fig.~\ref{fig:algo}(d)).
In DFT the polaron formation energy \EPOLDFT\ is defined as the total energy difference between the polaronic solution ($E_\mathrm{loc}^\mathrm{dist}$, with the excess charges localized in a locally distorted lattice sites) and the delocalized solution ($E_\mathrm{deloc}^\mathrm{undist}$, with all excess electrons uniformly delocalized over the lattice), ${E}_{\rm pol}^{\rm DFT}=E_\mathrm{loc}^\mathrm{dist} - E_\mathrm{deloc}^\mathrm{undist}$~\cite{Reticcioli2019}.
In our approach, the kernel regressor assigns a virtual single-polaron energy to every polaron in the configuration;
the predicted single polaron energies are summed up to return the total polaronic energy of the test configuration, which can be compared to the \EPOL\ calculated by DFT.

This scheme is used to first train the kernel regressor on an energy dataset of different polaronic configurations calculated by DFT.
Then, the trained ML algorithm can be applied to polaron configurations not included in the DFT training database, in order to drastically expand the exploration of the configuration space.
To ensure the quality of the predictions, the optimal polaron patterns identified by the ML algorithm can be ultimately tested and refined by few final DFT calculations.

In the following, we assess the quality and general efficiency of this computational strategy
on rutile \TO(110) and perovskite \STO(001) surfaces by studying the relative stability of millions of possible polaron configurations with varying polaron concentration (\ie\ various polaron densities), from the dilute to dense polaron limits.

\subsection{Rutile \TO(110)}

    \begin{figure}
        \centering
        \includegraphics[width=\textwidth]{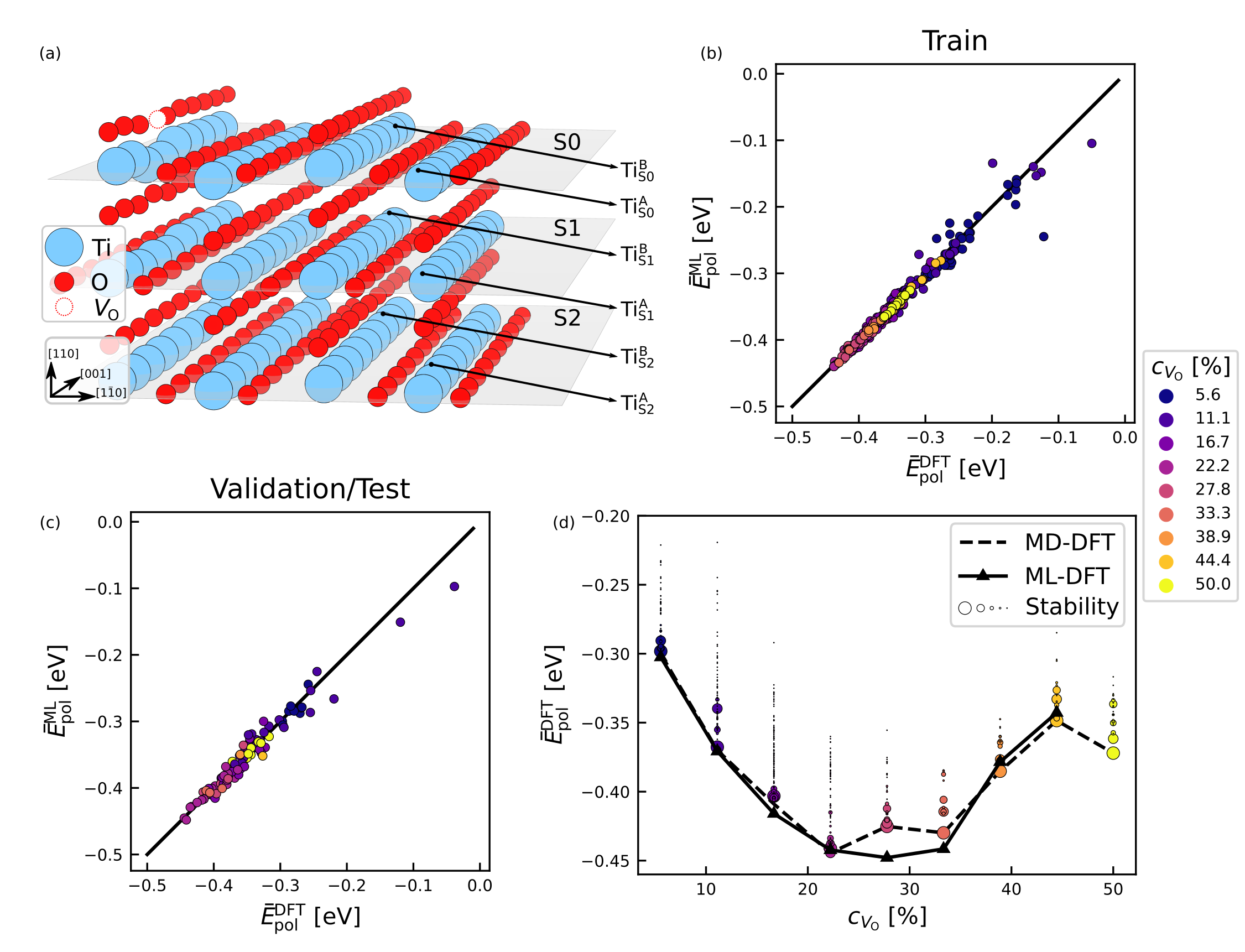}
        \caption{(a) Structural model of rutile \TO(110) with the correpsonding labeling. (b-c) Comparison of DFT and ML-derived mean polaron energies \barEPOLDFT and \barEPOLML for the training data in a randomized split are shown in (b) and the validation and test data are shown in (c).
        The color encodes the corresponding defect concentration of each configuration. (d) Collection of the results of the nine omitted defect concentrations (validation of the extrapolation capabilities) and the exhaustive search are shown. The black dotted curve gives the convex hull of polaronic energies from the training database and dots show \barEPOLDFT\ for each configuration, where the size of the dots corresponds to the favorability of each configuration as assigned by the ML model. The convex hull from the exhaustive exploration at the DFT-level is given in solid black with triangles indicating the most stable configuration's energy.} 
        \label{fig:res_tio2}
    \end{figure}

The structural unit of rutile \TO(110) is sketched in Fig.~\ref{fig:res_tio2}(a). The surface layer $S0$ consists of alternating rows of five- and sixfold-coordinated Ti atoms, referred to as Ti$^A$ and  Ti$^B$, respectively. Sub-surface and sub-sub-surface layers are indicated with the label $S1$ and $S2$, respectively.
Ti atoms coordinated below the surface \tisa / \tisb~sites are labelled as $A$- / $B$-sites, respectively, with an additional subscript $l$ indicating the layer, \eg\ \tisia\ refers to a Ti$^A$ sites located in the subsurface layer $S1$, whereas \tisiib\ indicates a B-type Ti located in $S2$.
The surface terminates with two parallel rows of oxygen atoms oriented along the [001] direction, which are easily removable to form surface oxygen vacancies (\vo).
Each (\vo) effectively donates two electrons that can be trapped in titanium sites converting two pristine Ti$^{4+}$ ions into Ti$^{3+}$ ions. Increasing the oxygen vacancy concentration \cvo\ leads to higher densities of polarons. Due to the combinatorial growth of accessible polaron configurations with progressively larger amounts of polarons, finding the most favorable distribution of polarons is a very challenging task.

Here we employ a 5 layers thick slab with a large 9$\times$2 two-dimensional (2D) unit cell containing 36 Ti sites per layer.~\cite{PhysRevX.7.031053,PhysRevB.98.045306}
Since the bottom two layers are fixed to bulk form (polaron inactive since no structural relaxation are allowed), this setup results in 108 potential trapping sites. We have inspected nine different defect concentrations starting from 1 \vo\ per 2D unit cell (corresponding to \cvo=5.5\%) up to 9 \vo\ (\cvo=50\%). Since polaron trapping at Ti sites follows a binomial distribution, the number of possible configurations ranges from $\approx 5\times 10^3$ (2 polarons, \cvo=5.5\%) to $\approx 10^{20}$ (18 polarons, \cvo=50\%), without considering symmetries (see Section S1.1 of the SM for a more detailed discussion).

To tackle this formidable problem, we rely on a previously generated MD polaron energy dataset~\cite{PhysRevX.7.031053,PhysRevB.98.045306}. 
It consists of 492 symmetrically-inequivalent polaron configurations, obtained by running MD simulations at nine different \VO\ concentration levels (ranging from \cvo$=5.5$\% up to \cvo$=50$\%). Details on the database are provided in Table ST1 of the SM.
The MD-based dataset suggest that at low \cvo, polarons preferably localize at \tisia\ sites, while at higher defect and polaron concentration the $S0$ sites become comparatively more populated~\cite{PhysRevB.98.045306,PhysRevX.7.031053}.

To begin with, this DFT-MD database was split in train, validation, and test datasets (70\%, 20\%, 10\%, respectively).
The ML scatter plots shown in  Figs.~\ref{fig:res_tio2} (b) and (c) show that the model can well reproduce the \barEPOLDFT\ with a mean square error MSE$\approx 1.2\cdot10^{-4}$ for training, and MSE$\approx 1.3\cdot10^{-4}$ for both validation and test (see Table ST5 in SM).
We note that configurations at very shallow \barEPOLDFT\ (\ie\ values higher than -0.2 eV) show larger errors, which we attribute to the lack of samples in the training dataset, owing to rare occurrence of these unfavorable polaronic solutions in the MD.
In fact, these polaron energies correspond to configurations where a polaron is localized on the unstable \tisib\ site.
Since we aim to identify the optimal polaron configuration, this issue is not of great concern, but it highlights the need of a well representative sampling of the polaron configurational space. As we show in Section~\ref{sec:sto},
a random sampling can overcome these limits of an MD-generated database.

To assess the validity of our method to predict configurations that strongly differ from the available training data, we have performed an additional test on unexplored polaron \vo\ concentrations, by adopting the ``omitted defect concentration'' strategy described below.
We constructed nine distinct test cases, one for each \vo\ concentration level available in the MD database.
In each of these cases, we limited the training dataset to samples belonging to only 8 of the 9 available defect concentrations. Finally, we tested the model on the remaining concentration, not used in the training.
We analyzed qualitative and quantitative aspects of this procedure by conducting a direct comparison between ML-derived and target DFT energies for specific defect concentrations.
We also compared the most stable configurations found via ML and via DFT-MD, by means of a convex hull diagram (reporting the most stable configuration at every defect concentration), 
to inspect whether ML is capable to deliver a qualitatively consistent energy landscape.
This is particularly important for the case under study, since experimental measurements indicate that, for \cvo$\approx 20\%$, highly defective \TO(110) undergoes a $1\times{1}$ to $1\times$2 surface reconstruction, which can be connected to a lowering of the polaron energy with increasing \cvo beyond a certain critical polaron concentration~\cite{ONISHI1994L783,PhysRevX.7.031053}.

The results are collected in Fig.~\ref{fig:res_tio2}(d) and in the Figure SF5 and Table ST5 of the SM.
Even though we notice an apparent loss of precision for configurations that are different from those included in the training data, leading to larger systematic errors (the resulting MSE of \barEPOL\ is of the order of 10$^{-3}$ for all test concentrations
), the ML model is still able to correctly identify the optimal arrangements of polarons from the MD dataset and find the correct minimum at \cvo =22\%. 
This is shown by the dot sizes in Fig.~\ref{fig:res_tio2}(d), indicating the ML-predicted favorability of a configuration relative to other predicted configurations in the respective defect concentration.
This result demonstrates that the DFT+ML scheme, trained with the available DFT dataset, is capable to predict with good accuracy the relative stability of solutions corresponding to different concentrations.

From these results, based on a ML-processing of the DFT-MD input database, we can conclude that a large-scale ML-based systematic search of new configurations (\ie~configurations not included in the original DFT database) should be feasible, provided that the final results are validated by a subsequent comparative DFT simulation.
In this way, an accelerated ML exploration of the full configurational space would be possible, and only a small fraction of first principles calculations would be necessary, limited to the most favorable configurations found by ML. 
Different methods could be used to perform this exhaustive search in the configuration space, and find the optimal polaron arrangement for each concentration: Our designed ``exhaustive search'' strategy is described in the following.

\subsubsection{Systematic search of configurations via ML}
An efficient search in the entire configuration space needs to address the problem of the combinatorial growth of polaron configurations with increasing polaron density (increasing \cvo).
In our DFT calculations, the unit cell contains 108 possible Ti trapping site, leading to $\binom{108}{18}\approx10^{21}$ possible arrangements of polarons in the most defective case (\cvo$=50\%$, 9~\VO\ defects in our cell with 18 polarons).
To cope with this huge combinatorial space, and avoid an inefficient blind search, we have developed a bottom-up searching strategy based on the concept of polaron building units. 
We have first trained our ML model on the entire DFT-MD database (492 inequivalent configurations).
We have then addressed the problem of two polarons (lowest considered concentration, \cvo$=5.5\%$) and predicted all $\binom{108}{2}=5778$ possible variants using the ML model, thus extending considerably the available DFT-MD database at that concentration.
From these 5778 configurations we selected the 100 with the most favorable polaron energy \EPOLML\ as calculated by ML, and used them as polaron building units to construct the polaron configurations at the next concentration level (\cvo$=11.1\%$, 4 polarons).
To this aim, we followed the same additive strategy, namely adding two additional polarons to each of the 100 two-polarons configurations, scrutinizing with the ML model all 106 not yet occupied hosting sites.
Similarly to the previous concentration, also in this case only a subset of 100 distinct, energetically most favorable configurations has been used to build the next configurations (\cvo$=16.7\%$, 6 polarons). The same protocol has been adopted for all higher concentrations.
Finally, at each concentration the three best ML-predicted polaron configurations have been verified by performing a DFT run using ML-predicted pattern as input with a selective initialization of the occupation matrix (see Section \ref{meth:dft} and Section S2 in the SM).
We refer to these final set of energies as exhaustive search ML-DFT energies.

The scheme described above allows for a quick calculation of roughly $4\cdot10^6$ configurational energies, excluding a larger number of highly unfavorable configurations, and keeping the computational cost constant among each defect concentration.
The results are displayed in Fig.~\ref{fig:res_tio2}(d), where we show the comparison between the reference MD-DFT convex hull (input database, dashed line) and the best obtained ML-DFT energies at each concentration (full lines). In Figure SF6 and SF8(a,b) of the SM most stable configurations and the distribution of polarons to specific sites, respectively, are shown.
The overall outcome is very satisfactory with a few positive aspects to note.
First, the ML-aided DFT procedure finds a lowest absolute minimum not present in the MD database, clearly demonstrating the remarkable efficiency of the proposed ML scheme in exploring the configurational space.
The absolute minimum is shifted from \cvo = 22.2\% to 27.8\% resulting in a much smoother shape of the ML-derived convex hull.
At defect concentrations lower than 22\% the two curves essentially overlap, with the ML configurations all slightly lower than the best MD configurations (in this case with a modest improvement of few meV).
More subtle are cases at \cvo$>33\%$, where polarons start populating Ti$^B_{S0}$-sites.
This is due to a poor sampling of Ti$^B_{S0}$-site in the MD simulations, and to a tendency at high concentration towards charge delocalization which impedes a site-specific assignment of the excess charge, pushing the system out of the polaron regime.
As already mentioned, to avoid an unrealistic densely-packed arrangement of polarons or occupation of highly unfavorable trapping sites (unlikely to be observed in experiment~\cite{Kruger2008}), TiO$_2$(110) samples undergo a polaron-driven (1$\times$1) $\rightarrow$ (1$\times$2) structural reconstruction at \cvo $\approx$ 20\%~\cite{PhysRevX.7.031053}.
Therefore, the apparent inefficiency of the ML search above the critical concentration is not an intrinsic deficiency of the designed strategy, rather it should be traced back to physical arguments that hamper the construction of a suitable database for an unrealistic situation.
Forcing the structure to maintain the (1$\times$1) symmetry leads to formation of \tisb\ polarons, which are energetically not favorable, and tend towards charge delocalization and to polaron diffusion to the bulk (trapping at S$_2$ sites)~\cite{Shibuya2017b}.
In this sense the present ML data provide further support and validation to the DFT-based description of the polaron-driven surface reconstruction discussed in Ref.~\citeonline{PhysRevX.7.031053}.

\subsection{SrTiO$_3$(001)}
\label{sec:sto}

    \begin{figure}
        \centering
        \includegraphics[width=\textwidth]{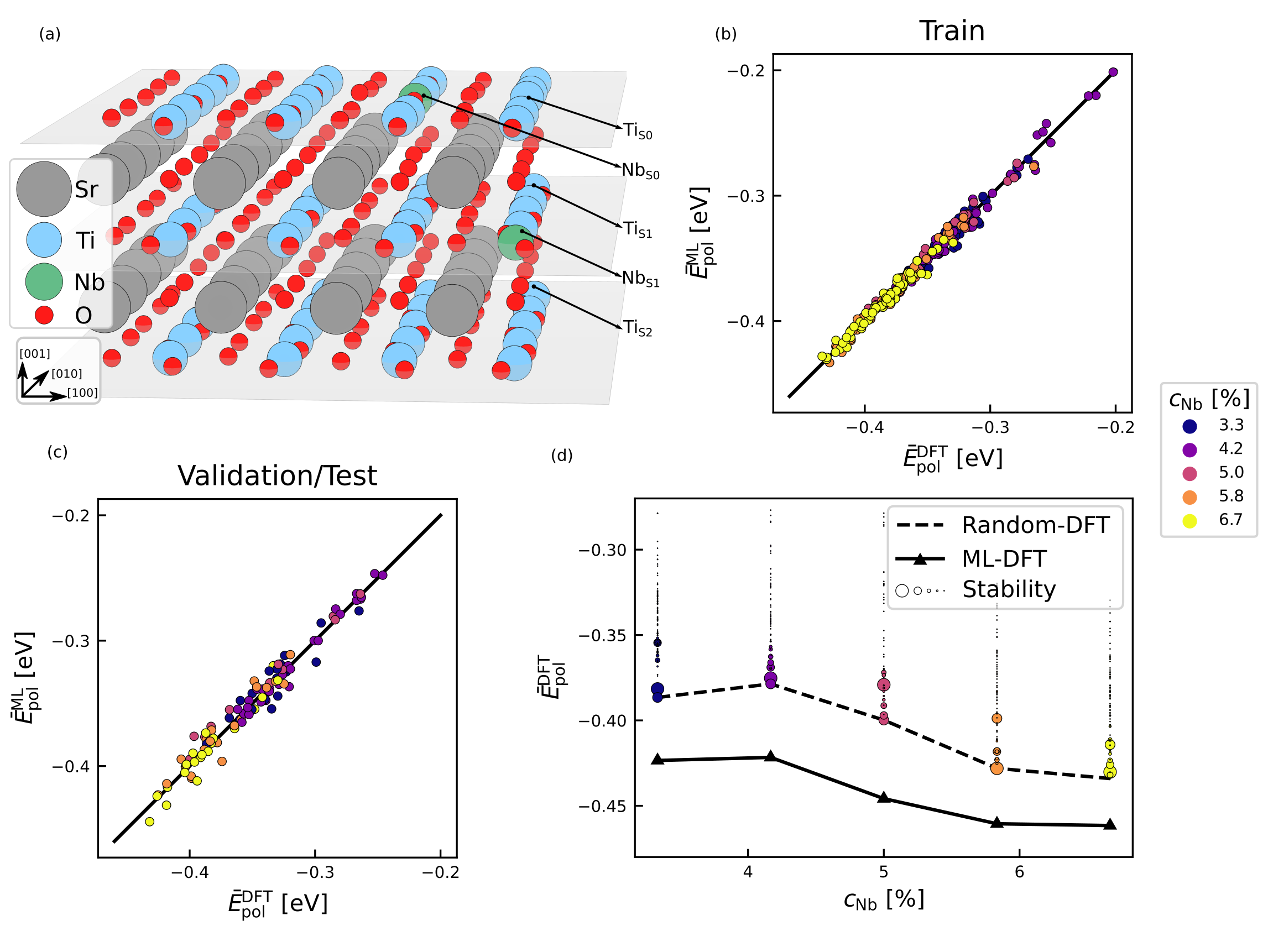}
        \caption{Collection of results when applying the methodology to SrTiO$_3$(001). (a) Model of the surface slab and the corresponding labeling. Comparison of \barEPOLDFT\ and \barEPOLML\ for the training data in a randomized split are collected in (b) and the validation and test data are shown in (c). The color encodes the corresponding Nb-dopant concentration (\cnb) of each configuration. (d) Collection of the results of the five omitted defect concentration cases and the exhaustive search are shown. The black dotted curve gives the hull of polaronic energies obtained from the training database and dots show \barEPOLDFT for each configuration, where the size of the dots corresponds to the ML-model's assigned favorability of each configuration. The Nb-concentration dependent polaronic ground state energy from the exhaustive exploration is given solid black with triangles marking the ML-predicted configuration's energy (all energies at DFT-level).}
        \label{fig:res_srtio3}
    \end{figure}
    
    To assess the degree of transferability and generality of the proposed methodology we applied a similar scheme to Nb-doped~\STO(001), a different structure (atomically flat perovskite surface, see Figure~\ref{fig:res_srtio3}(a)) with different source of excess charge (chemical doping instead of \vo), resulting in different type of interaction categories (see Section~\ref{method:desc}).
    To further generalize our DFT+ML procedure we decided to follow a different strategy to build the necessary DFT database. Instead of MD runs, which require long execution time and could result in an inefficient exploration of the configuration ground state, we have adopted a randomly generated polaronic database using the occupation matrix approach (see Table ST2, Section ~\ref{meth:dft} and Section S2 in the SM).
    This procedure has few advantages as compared to the MD-sampling methodology. First of all, obtained samples are less correlated than in the MD-runs. Secondly, the bias towards low energy configurations, which are disproportionately more often visited in an MD-simulation, is removed. This should result in a more general model that has higher accuracy across all possible polaron patterns. Lastly, and most importantly, it fully bypasses the cost of the MD-simulations and only the structural relaxations for each distinct polaronic pattern has to be performed (see Section S5 in the SM).
    Following the same protocol as discussed in the previous section, we again performed a randomized split for training the model on the entire database, consisting of 379 polaronic configurations.
    Following this we assessed the extrapolation capabilities by testing the model on defect concentrations not present in the training data. Lastly, we performed an exhaustive bottom-up search for all defect concentrations, were we then evaluated most favorable predictions at the DFT-level.
    
    Figure~\ref{fig:res_srtio3} (b) and (c) show the results from the randomized data-split. The model is converged to a similar extent as in the case of \TO\ (\barEPOL\ MSE of $5.7\cdot10^{-5}$, $9.3\cdot10^{-5}$ for training and test data, respectively), and,  
    noticeable and unlike~\TO, the model delivers consistent accuracy at virtually all energies. This increased performance most likely originates from the less biased energy database obtained by the random sampling. In the case of omitted defect concentrations (relative favorablity prediction shown in Figure \ref{fig:res_srtio3} (d)), the model correctly extrapolates the low energy configurations for the omitted concentration based on the energies of the 4 included concentrations, and 
    suffers to a smaller extent from systematic errors as compared to \TO\ (compare \TO\ and \STO\ cases in Tables~ST5 and ST6 as well as Figure SF5).
    
    Finally, we tested the efficiency of the exhaustive ML search, by exploring 2.25$\cdot10^6$ nonequivalent configurations.
    The outcome (see dotted and full line in Figure~\ref{fig:res_srtio3}(d)), clearly shows that this ML-augmented scheme outperforms the standard random-database approach, as it finds 
    multiple configurations lower in energy than the minima identified in the randomized search, at any polaron concentration.
    We confirmed the five lowest energies \EPOLML\ predicted by the ML exhaustive search at each Nb-concentration by a comparative static DFT-calculation and each of the ML-predicted polaron pattern was found more stable than the optimal pattern included in the training database.
    The most stable configuration predicted by ML (see Figure SF7 for a collection of the most optimal polaron configurations) typically improved the mean polaronic energy \barEPOL\ by 30 to 50~meV compared to the reference data. 
    Interestingly, the results of the exhaustive ML search suggest a rationale for the most stable configuration based on a few simple rules: the energetically most stable configurations host polarons in the surface and subsurface layer, usually placed as close as possible to the Nb-dopants (preferably below or above a dopant rather than in the same atomic layer (see site distribution of favorable configurations collected in Figure SF8(c) in the SM).
    
\section{Discussion and Conclusions}
    In this paper we presented an ML-aided procedure to enhance and accelerate the identification of small polaron ground state configurations in multi-defect systems with varying defect concentration, by employing simple and general descriptors based on distance-dependent interaction categories and a standard kernel regression fed by a DFT energy database. 
    We tested and discussed a few alternative protocols:
    
    (i) A conventional train/validation/test ML protocol.
    
    (ii) Omitted-defect concentration model, based on extrapolating the polaron energy for a given defect concentration from the DFT energies obtained at other concentrations.
    
    (iii) Exhaustive ML search. An exploration of the polaron configurational space based on a guided bottom-up selection of the most favorable configuration from all possible nonequivalent configurations at each given defect concentration.
    
    We assessed the generality of the procedure by applying it to two different materials (\TO(110) and \STO(001)) with different types of defects (\vo\ and Nb dopants) and adopting different strategies to construct the DFT database (MD and random sampling). 
    Our data indicates that a randomized sampling approach is superior to MD-generated database which suffers from undesirable correlation between the MD-generated configurations and excessive computational cost. Importantly, the combination of random sampling and exhaustive ML search results in a robust algorithm that delivers very good results, as demonstrated for \STO(001) where this procedure leads to a systematic improvement at all explored concentration: the exhaustive ML search finds configurations with lower 0-K DFT energy as compared to those included in the input database.
    
    While our model has been applied to the identification of polaron configurations with static dopant/vacancy patterns, it can be further extended to consider 
    optimized configurations with mobile point defects considering other type of defects (e.g. hydrogen adatom or Ti interstitials) and other materials~\cite{PhysRevLett.122.096101}.
    In fact, the descriptor only relies on identifying polaron hosting sites with different local coordination and lattice symmetry and their relative position with respect to the surface, to structure a list of distances between polarons and defects. 
    From this information, the descriptor structure can easily be attained for any material and only few parameters (\ie\ number of included distances in each interaction category and cutoff radius) need to be determined to optimize the performance. 
    
    A final positive aspect offered by the proposed method is the arbitrary scalability with respect to the supercell size, enabling access to large scale simulations,
    where defect arrangements could be precisely aligned with experimental data, to determine likely polaronic configurations
    observed in surface imaging techniques. Also, qualitative extrapolations and interpolations to defect concentrations where no data is available and predictions seem plausible in the presented test cases and could be further developed.

\section{Methods} \label{sec:method}
    \subsection{Density Functional Theory} \label{meth:dft}
        Density-functional theory (DFT) and first-principles molecular dynamics (MD) calculations were performed using the Vienna Ab-initio Simulation Package (VASP)~\cite{kresse_efficiency_1996, kresse_efficient_1996}.
        For our DFT+$U$ calculations we adopted the generalized gradient approximation with Perdew, Burke, and Ernzerhof parametrization (PBE)~\cite{PhysRevLett.77.3865}, including an on-site effective $U=3.9$~eV enacted on the d-orbitals of Ti atoms in the case of rutile \TO\ (previously determined by constrained random-phase approximation (cRPA)~\cite{PhysRevLett.113.086402}) and $U=4.5$~eV for \STO, here enacted on the d-orbitals of Ti and Nb, in line with the cRPA value determined in previous works.\cite{PhysRevB.91.085204, Hou2010, Choi2013}
        We used the $\Gamma$ point only for the integration in the reciprocal space, and standard convergence criteria with a plane-wave energy cutoff of 250~eV~\cite{PhysRevB.98.045306} for rutile \TO, and 350~eV for \STO.
        
        The rutile \TO(110) and \STO(001) surfaces were modeled by super cells, containing five stochiometric layers in large two-dimensional $9\times2$ and $6\times4$ unit cells, respectively. The three surface layers and the corresponding labeling are shown in Figure ~\ref{fig:res_tio2}(a) and Figure ~\ref{fig:res_srtio3}(a). In both cases the bottom two stochiometric layers were kept fixed at bulk positions in order to mimic and retrieve the electronic and structural properties of the bulk. Oxygen vacancies (\vo) on the two-fold coordinated O sites on the \TO\ surface, and Nb dopants replacing Ti atoms on the surface and subsurface \STO\ layers, were modeled at nine and five different concentration levels, \cvo\ and \cnb\ respectively.
        The defect positions were chosen such that inter-defect distances are maximized,
        and the concentrations are given in percentage with respect to the number of two-fold surface oxygen sites in \TO\ and the total number of Ti sites in \STO.
        \STO\ is known to exhibit a wide variety of surface reconstructions, and it has recently been shown that flat bulk-terminated (001)-surface with surface defects can be stabilized using novel cleaving procedure~\cite{sto1,sto2}.
        
        The polaronic localization sites were identified by inspecting the size of the local magnetic moments on Ti ions (larger than $0.5~\mu_{\rm B}$) and relaxations to 0 K have been performed from each distinct polaron localization pattern.
        In the case of rutile \TO\ nonequivalent polaron configurations were generated via MD at high temperature (700~K). 
        Instead, to build a database of polaronic configurations in ~\STO, for each concentration, an appropriate number of Ti sites within the top three layers were randomly chosen to host the polarons. For localizing the excess charge in the selected Ti hosting site we employed the occupation matrix control tool~\cite{occupation_matrix_control}. This tool allows us to constrain the electron density matrix of atoms in the cell directly, such that a polaron can be placed explicitly at the desired site, and in the desired orbital. Polaron configurations suggested by the exhaustive ML-aided search were always initialized via occupation matrix control, starting from initial occupation matrices taken from the training databases. Examples of occupation matrices are collected in the Supplementary Material in Section S2.
        
        A final note on the calculation of the DFT polaron energy \EPOLDFT: DFT calculations do not grant access to formation energies of individual polarons, but only to the total energy of all polarons in the unit cell. In order to compare polaron formation energies for configurations with different polaron concentration, we defined a mean polaronic energy $\bar{E}_{\rm pol}^{\rm DFT} = E_{\rm pol}^{\rm DFT}/N_{\rm pol}$, obtained by averaging over the number of polarons $N_{\rm pol}$ in a given configuration. This operation allows us to compare the polaron energies obtained for different defect concetrations.
        
    \subsection{Descriptors} \label{method:desc}

To design a suitable descriptor, we have exploited the fact that polaron energies are affected by two main interactions~\cite{PhysRevB.98.045306}:

(i) Electrostatic interaction between charged defects: attractive interaction between negatively charged polarons and positively 
charged defects and polaron-polaron repulsion, both rapidly decreases with increasing spatial separation.

(ii) Polaron orbital symmetry, determined by the local coordination and symmetry of the hosting site as well as by the distance 
from the surface layer (For example, in  rutile \TO(110) the distance-dependence of the polaron-polaron \tisia-\tisia\ interaction is different from \tisia-\tisa~\cite{PhysRevB.98.045306}).

Based on these principles, we designed a descriptor composed of a list of pairwise polaron-polaron and polaron-defect interactions, defined by the corresponding inter-polaron and polaron-defect distances within a cutoff-sphere $R_c$ around each polaron (defect: \vo\ and Nb, for TiO$_2$ and SrTiO$_3$, respectively).
In this way, fixed portions of the descriptor vector can be assigned to specific interaction categories. An interaction category always depends on the local coordination of the hosting site (\eg\ \tisia) and the local coordination of its interacting partner (\eg\ \tisa). Within each interaction category, a fixed number of preassigned slots $n$ can be used, containing rescaled distances according to the expression: 

        \begin{equation} \label{eq:rescale}
            f_c(d)=
            \begin{cases}
            \frac{1}{2}\left(1+\cos{\frac{\pi d}{R_c}}\right)\quad &\text{if}~d\leq R_c\\
            0\quad &\text{else}
            \end{cases}
        \end{equation}

The rescaling allows that the vector is filled with zeros, were no interaction is present. For an exemplary description of the evaluation of a polaron descriptor see Section S3 in the SM. Mind that each descriptor corresponds to the environment of a single polaron. Therefore, the number of descriptors from the databases are 4368 and 2257 for \TO\ and \STO, respectively.

We complete this section by providing a brief description of the resulting interaction categories for \TO\ and \STO. More details can be found in Section S3 in the Supplementary Materials.
        
\subsubsection{Interaction categories in \TO(110)}
In rutile-\TO\ the three topmost layers contain hosting sites with two different local orientations, for a total of 36 interaction categories (Ti$_{\rm Si}^{\rm X}$-Ti$_{\rm Sj}^{\rm Y}$, with X,Y $\in \{\rm A,\rm B\}$ and i,j $\in \{0,1,2\}$). Due to the employed $9\times2$ supercell and the slightly different distance-dependence of stacked and non-stacked hosting sites of same local coordination~\cite{PhysRevB.98.045306}, we distinguish between stacked ($\rm A$- or $\rm B$-sites) and non stacked ($\rm A'$- or $\rm B'$-sites). This adds 18 additional interaction categories.
Lastly, a category for the interaction with the \vo\ is added for each differently coordinated sites (Ti$_{\rm Si}^{\rm X}$-\vo, with X $\in \{\rm A,\rm B\}$ and i $\in \{0,1,2\}$), resulting in 60 possible interaction categories. Three shortest distances per interaction category and a cutoff radius of 15~\AA\ resulted in optimal model predictions for this descriptor. The full list of interaction categories for \TO\ is given in the Supplementary Materials in Table~ST3 and examples of interaction categories are collected in Figures~SF1 and SF2.

\subsubsection{Interaction categories \STO(001)}
Cubic and atomically flat \STO\ has a more symmetrical structure, which results in a simpler descriptor. The three topmost layers contain identically coordinated hosting sites, leading to nine different interaction categories (Ti$_{\rm Si}$-Ti$_{\rm Sj}$, with i,j $\in \{0,1,2\}$). However, a more fine grained distinction of polaron-dopant interactions is necessary, since dopants can in principle lay in any layer (unilke \vo\ in rutile \TO(110)). Therefore, we consider six additional interaction categories (Ti$_{\rm Si}$-Nb$_{\rm Sj}$, with i $\in \{0,1,2\}$ and j $\in \{0,1\}$), resulting in a total of 15 different interaction categories with an optimal number of four included distances per category and a $R_c=13$~\AA. A full list of interaction categories for \STO\ is given in the Supplementary Materials in Table ST4 and examples of interaction categories are collected in Figure SF1.
        
\vspace{4mm}

    \subsection{Machine learning model} \label{method:ml}
        We have used instance-based learning in form of kernel regression \cite{bishop_pattern_2006}, where a kernel-function $k(\cdot,\cdot)$ (or similarity measure) of the descriptor of interest $x'$ and all descriptors $x_i$ of the training set is calculated and scalar multiplied with the optimized regression parameters $\alpha$, giving a weighted mean of the target quantity based on similarity to training instances: 
            \begin{equation}
                \label{eq:krr}
                    y' = \sum\limits_{i=1}^N k(x_i,x') \alpha_i
            \end{equation}
        We determine optimal regression parameters of each kernel regressor, corresponding to a specific type of polaron, via backpropagation and gradient-descent performed on the sum of each kernel regressors prediction. We found that training the regressors with a stochastic gradient descent varient, results in better extrapolation capabilities then performing an exact fit. To optimize regression parameters of the different kernel regressors on the training data, the predicted energy \EPOLML\ is used to calculate the loss function with respect to the target polaron energy \EPOLDFT\ of the configuration from the training dataset .
        We employed an adapted mean squared error loss function $J$, where the error of each training sample is normalized to the number of polarons in the configuration ($N_{\rm{pol},i}$).
                \begin{equation}
                \label{eq:loss}
                    J(E_{\rm pol}^{\rm ML},E_{\rm pol}^{\rm DFT})= \sum\limits_{i} (E_{\rm{pol},i}^{\rm ML}-E_{\rm{pol},i}^{\rm DFT})^2/N_{\rm{pol},i}
                \end{equation}
        Without a normalization of the error to the number of polarons, the model tended to systematically underpredict energies at low defect concentrations, which can likely be attributed to a greater total number of polarons at high defect concentrations.
        More polarons result in a cumulative higher error and the model shifts towards providing a better fit for configurations with many polarons at high defect concentrations -- compromising accuracy for configurations with fewer polarons.
        For the optimization of the loss function, an ADAM optimizer\cite{kingma_adam_2014} has been used with a learning rate set to 0.0001, which resulted in slow yet consistent convergence of regression parameters (see Figure SF3 in the SM).
        It has been found that an initialization of regression parameters set to 0 resulted in a faster convergence than a random initialization.
        A batch size of 64 randomly chosen configurations per epoch, and a weight decay of 0.1 to enforce small regression parameters, also benefited convergence of the model to consistent performance throughout many tests.
        We used the Laplacian kernel
        \begin{equation}
        \label{eq:laplacian}
                    k(x_i,x_j) = \exp{\left(-\gamma~||x_i-x_j||_1\right)}
        \end{equation}
        for all models and found that a hyperparameter $\gamma$ between 0.1 and 0.5 leads to optimal results (see Figure SF4 in the SM).
        The algorithm has been implementd using NumPy\cite{van_der_walt_numpy_2011} and Scikit-Learn\cite{pedregosa_scikit-learn_2011} for preprocessing, and PyTorch\cite{NEURIPS2019_9015} to perform the backpropagation and optimization of parameters. 

\section*{Data availability}
    The data and code used in the presented article is available from the authors upon reasonable request.
\bibliography{references}

\noindent\textbf{Acknowledgements}\\
This work was supported by the Austrian Science
Fund (FWF) project POLOX (Grant No. I 2460-N36), project Super (Grant No. P 32148-N36) and the SFB-F81 project TACO. The computational results have been achieved using the Vienna Scientific Cluster (VSC).

\noindent\textbf{Author contributions}\\
CF conceptualized the research. VB has constructed and extended the model and executed all calculations on \TO\ and FE performed all calculations on \STO, supervised by MR \& CF. CF \& VB wrote the first draft. All authors contributed to a substantial discussion of the results and a critical reading of the manuscript.

\noindent\textbf{Competing interests}\\
The authors declare no competing interests. \\

\end{document}


\maketitle

\tableofcontents

\newpage

\section{Datasets}
    The distribution of polarons in the training databases for different \vo\ and Nb-doping concentrations are collected in Table \ref{tab:dataset_to} and \ref{tab:dataset_sto}, respectively.
    \begin{table}[h!]
        \centering
        \begin{tabular}{c | c | c | c | c | c | c | c }
        $c_{V_O}$ & \# Configurations & \tisa & \tisb & \tisia & \tisib & \tisiia & \tisiib\\
        \hline \hline
        5.5\%  & 52  & 6   & 0   & 97  & 1  & 0  & 0 \\
        11.1\% & 86  & 99  & 0   & 229 & 13 & 3  & 0 \\
        16.7\% & 116 & 214 & 0   & 462 & 18 & 2  & 0 \\
        22.2\% & 61  & 192 & 0   & 239 & 39 & 18 & 0 \\
        27.8\% & 27  & 125 & 0   & 108 & 28 & 9  & 0 \\
        33.3\% & 11  & 44  & 16  & 56  & 16 & 0  & 0 \\
        38.9\% & 19  & 98  & 56  & 95  & 11 & 6  & 0 \\
        44.4\% & 46  & 219 & 170 & 261 & 57 & 9  & 20 \\
        50\%   & 74  & 450 & 383 & 338 & 97 & 64 & 0 \\
        \hline \hline
        total & 492 & 1447 & 625 & 1885 & 280 & 111 & 20 \\
       \end{tabular}
     \caption{FPMD dataset of configurations in rutile \TO(110). Number of inequivalent polaronic configurations generated in FPMD for every defect concentration, and corresponding site-dependent polaron occupation (number of times the site is occupied by a polaron). 
     }
     \label{tab:dataset_to}
    \end{table}
        
    \begin{table}[h!]
        \centering
        \begin{tabular}{c | c | c | c | c }
        $c_{\rm Nb}$ & \# Configurations & Ti$_{\rm S0}$ & Ti$_{\rm S1}$ & Ti$_{\rm S2}$ \\
        \hline \hline
        3.3\% & 95  & 146 & 174 & 60 \\
        4.2\% & 79  & 153 & 73  & 169 \\
        5.0\% & 42  & 102 & 50  & 100 \\
        5.8\% & 74  & 178 & 185 & 155 \\
        6.7\% & 89  & 254 & 236 & 222 \\

        \hline \hline
        total & 379 & 833 & 718 & 706 \\
       \end{tabular}
     \caption{Randomly generated dataset of configurations in \STO(001). Number of inequivalent polaronic configurations generated for every defect concentration, and corresponding site-dependent polaron occupation (number of times the site is occupied by a polaron).
     }
     \label{tab:dataset_sto}
    \end{table}

\subsection{Number of possible configurations}
    We briefly discuss the number of possible configurations in the \TO\ super cell.
    In our setup, consisting of a $9\times2$ large unit-cell with two of the five stochiometric layers fixed to bulk positions (see Methods in the main text), the \TO\ slab contains 108 Ti sites that can possibly host polarons (36 sites per $S0$, $S1$ and $S2$ layer).
    In the simplest case of one oxygen vacancy and two excess electrons in the slab (\cvo$=5.5\%$), the number of possible polaronic configurations (with no symmetry applied) is given by the binomial coefficient, ${108 \choose 2}=5778$. At higher concentration, the number of configurations explodes, \eg\ for 4 oxygen vacancies we obtain $108 \choose 8$ $\approx 3\cdot10^{11}$ polaronic configurations.
    We note that, especially at high defect concentration, exploiting symmetry operations to simplify the problem does not bring any considerable advantage.

\clearpage

\section{Occupation matrices}

We used the Occupation Matrix Control scheme as developed in Ref.~\cite{occupation_matrix_control} to localize polarons at specific sites in a two step protocol.
In the first step, occupation matrices of chosen Ti sites are fixed in order to ensure polaron localization at the specified sites.
The second step consists of an unrestricted relaxation, which starts from the previously determined distorted structure.
The second step is crucial to obtain reliable self-consistent results of the polaron trapping.

In the following, we list the spin-resolved density matrices used in the first step of the relaxation procedure. In the case of rutile \TO(110) we employed specific occupation matrices for every site type (\tisa, \tisia, \tisib, \tisb), here listed with the corresponding spin-channel $\uparrow$ and $\downarrow$ arrows. All occupation matrices have been extracted from a single polaron configuration at \cvo=38.9\%. Polarons at \tisa, \tisia, and \tisib\ showed well defined orbital characters, while the unstable \tisb\ polarons varied in their specific orbital character. Below, the used input occupation matrices are listed:

\[
\rm{Ti}_{\rm S0}^{\rm A},\uparrow=
\begin{bmatrix}
  0.32 &-0.01 &-0.03 &-0.01 & 0.01\\
 -0.01 & 0.57 &-0.06 &-0.38 &-0.19\\
 -0.03 &-0.06 & 0.19 & 0.04 &-0.01\\
 -0.01 &-0.38 & 0.04 & 0.40 & 0.15\\
  0.01 &-0.19 &-0.01 & 0.15 & 0.15
    
\end{bmatrix}
\;\rm{Ti}_{\rm S0}^{\rm A},\downarrow=
\begin{bmatrix}
  0.30 &-0.00 &-0.03 &-0.00 & 0.01\\
 -0.00 & 0.06 &-0.00 & 0.02 & 0.00\\
 -0.03 &-0.00 & 0.18 &-0.00 &-0.03\\
 -0.00 & 0.02 &-0.00 & 0.08 &-0.00\\
  0.01 & 0.00 &-0.03 &-0.00 & 0.08

\end{bmatrix} 
\]

\[
\rm{Ti}_{\rm S1}^{\rm A},\uparrow=
\begin{bmatrix}
  0.09 & 0.00 &-0.02 &-0.00 & 0.01\\
  0.00 & 0.09 &-0.00 &-0.00 &-0.00\\
 -0.02 &-0.00 & 0.76 & 0.00 &-0.23\\
 -0.00 &-0.00 & 0.00 & 0.28 &-0.00\\
  0.01 &-0.00 &-0.23 &-0.00 & 0.43
    
\end{bmatrix}
\;\rm{Ti}_{\rm S1}^{\rm A},\downarrow=
\begin{bmatrix}
  0.08 & 0.00 &-0.00 &-0.00 &-0.00\\
  0.00 & 0.08 &-0.00 &-0.00 &-0.00\\
 -0.00 &-0.00 & 0.10 &-0.00 & 0.09\\
 -0.00 &-0.00 &-0.00 & 0.26 & 0.00\\
 -0.00 &-0.00 & 0.09 & 0.00 & 0.24

\end{bmatrix} 
\]

\[
\rm{Ti}_{\rm S1}^{\rm B},\uparrow=
\begin{bmatrix}
 0.29	&-0.00	&-0.00 	& 0.00	&-0.01 \\	
-0.00	& 0.08	&-0.00	& 0.00	&-0.01 \\	
-0.00	&-0.00  & 0.29	& 0.00	& 0.06 \\
 0.00	& 0.00	& 0.00	& 0.09 	&-0.02 \\
-0.01	&-0.01	& 0.06	&-0.02	& 0.88

\end{bmatrix}
\;\rm{Ti}_{\rm S1}^{\rm B},\downarrow=
\begin{bmatrix}
 0.28	&-0.00 	&-0.00	& 0.00	& 0.00 \\	
-0.00	& 0.07	&-0.00	& 0.00	& 0.00 \\	
-0.00	&-0.00	& 0.26	& 0.00	&-0.02 \\	
 0.00	& 0.00	& 0.00	& 0.08	&-0.00 \\	
 0.00	& 0.00	&-0.02	&-0.00	& 0.06 	

\end{bmatrix} 
\]

\[
\rm{Ti}_{\rm S0}^{\rm B},\uparrow=
\begin{bmatrix}
0.11 & -0.00 & -0.05 & 0.03 & 0.05\\
-0.00 & 0.10 & -0.01 & 0.00 & 0.03\\
-0.05 & -0.01 & 0.58 & -0.21 & -0.24\\
0.03 & 0.00 & -0.21 & 0.35 & 0.09\\
0.05 & 0.03 & -0.24 & 0.09 & 0.49

\end{bmatrix}
\;\rm{Ti}_{\rm S0}^{\rm B},\downarrow=
\begin{bmatrix}
0.09 & -0.01 & 0.01 & 0.00 & 0.01\\
-0.01 & 0.08 & 0.00 & -0.00 & 0.01\\
0.01 & 0.00 & 0.11 & 0.01 & 0.08\\
0.00 & -0.00 & 0.01 & 0.24 & -0.06\\
0.01 & 0.01 & 0.08 & -0.06 & 0.25

\end{bmatrix} 
\]

\[
\rm{Ti}_{\rm S0}^{\rm B},\uparrow=
\begin{bmatrix}
0.54 & 0.41 & 0.03 & 0.00 & 0.02\\
0.41 & 0.48 & 0.03 & 0.00 & 0.02\\
0.03 & 0.03 & 0.13 & -0.03 & 0.03\\
0.00 & 0.00 & -0.03 & 0.32 & -0.06\\
0.02 & 0.02 & 0.03 & -0.06 & 0.24

\end{bmatrix}
\;\rm{Ti}_{\rm S0}^{\rm B},\downarrow=
\begin{bmatrix}
0.07 & -0.02 & -0.00 & -0.00 & -0.01\\
-0.02 & 0.06 & -0.00 & 0.00 & -0.01\\
-0.00 & -0.00 & 0.10 & -0.01 & 0.05\\
-0.00 & 0.00 & -0.01 & 0.26 & -0.08\\
-0.01 & -0.01 & 0.05 & -0.08 & 0.21

\end{bmatrix} 
\]

For \STO(001), we employed the same occupation matrix for all localization sites in both the generation of the reference database and in the exhaustive search:
\[
\rm{Ti},\uparrow=
\begin{bmatrix}
  0.28 &-0.00 &-0.02 & 0.00 &-0.07\\
 -0.00 & 0.41 &-0.00 &-0.40 & 0.00\\
 -0.02 &-0.00 & 0.24 & 0.00 & 0.00\\
  0.00 &-0.40 & 0.00 & 0.59 &-0.00\\
 -0.07 & 0.00 & 0.00 &-0.00 & 0.12
\end{bmatrix}
\;\rm{Ti},\downarrow=
\begin{bmatrix}
  0.27 & 0.00 &-0.02 & 0.00 &-0.07\\
  0.00 & 0.07 &-0.00 & 0.01 &-0.00\\
 -0.02 &-0.00 & 0.23 &-0.00 & 0.00\\
  0.00 & 0.01 &-0.00 & 0.06 & 0.00\\
 -0.07 &-0.00 & 0.00 & 0.00 & 0.11

\end{bmatrix} 
\]

\newpage

\section{Descriptors}
    In this section we briefly describe the most important concepts employed for constructing our descriptor.
\subsection{Interaction Categories} \label{sec:int_cat}

    \begin{figure}[h!]
        \centering
        \includegraphics[width=1.0\textwidth]{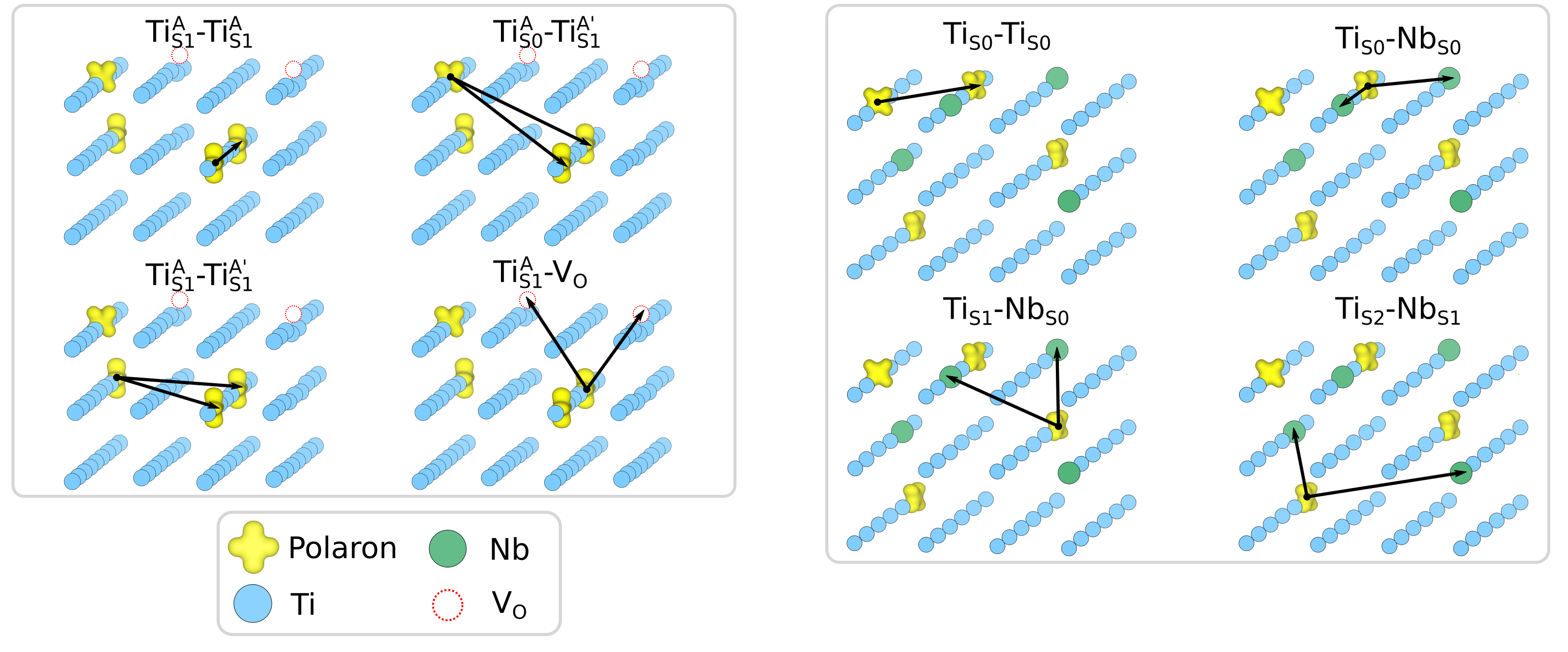}
        \caption{(left) Four of the possible 60 interaction categories in a rutile \TO(110) polaron configuration at \cvo$=11.1\%$ are shown. For clarity, O-atoms are hidden and the locations of oxygen vacancies are marked. The starting point of arrows indicates the polaron, whose descriptor ought to be calculated and the arrow indicates interactions with polarons/defects at specific sites (vectors connecting the considered sites are not necessarily the shortest distance in the periodic cell, although in practice the minimum image convention was used to construct descriptors). (right) Simlilarly to the left hand side, some examples of interaction categories in \STO(001) are depicted. Again, O- and Sr-atoms have been omitted for clarity, but Nb-dopants are displayed.}
        \label{fig:int_cat}
    \end{figure}
    
To ensure consistent representations of the polaron environment we structure the 2-body-interaction terms used to construct the descriptor via interaction categories.
An interaction category depends on the type of polaron-hosting site, and on the nature of interaction (whether with another polaron of with a donor defect).
Figure~\ref{fig:int_cat} collects examples of interaction categories for the two materials of our study.
The arrows in the left side of Figure~\ref{fig:int_cat} show four interaction categories for rutile \TO(110):
polaron-polaron on the same-row (\tisia-\tisia), non-stacked on different layers (\tisa-Ti$_{\rm S1}^{\rm A'}$), non-stacked on same layer (\tisia-Ti$_{\rm S1}^{\rm A'}$), and polaron-defect (\tisia-\vo).
The right panel similarly shows four possible interaction categories in \STO: polaron-polaron on the same layer (Ti$_{\rm S0}$-Ti$_{\rm S0}$), polaron-defect on the same (Ti$_{\rm S0}$-Nb$_{\rm S0}$) and different (Ti$_{\rm S1}$-Nb$_{\rm S0}$, Ti$_{\rm S2}$-Nb$_{\rm S1}$) layers.
    A full list of possible interaction categories for \TO(110) and \STO(001) is provided in Table~\ref{tab:1} and~\ref{tab:2}, respectively.

\begin{table}[h!] 

    \centering
    \small
    \setlength\tabcolsep{2pt}
    \begin{tabular}{c|l|llllll}
    Layers & \multicolumn{1}{c|}{Interaction type}                & \multicolumn{6}{c}{Trapping site}                                                                                                                                  \\ \hline
                     &   Ti$-$Ti or Ti$-$\vo                 & \tisa                 & \tisb                 & \tisia                 & \tisib                 & \tisiia                 & \tisiib                 \\ \hline
    \multirow{3}{*}{$\rm Sl- \rm S0$} & $\rm A-A$/$\rm B-B$ (stacked)                & \tisa-\tisa  & \tisb-\tisb  & \tisia-\tisa  & \tisib-\tisb  & \tisiia-\tisa  & \tisiib-\tisb  \\
                           & $\rm A-A'$/$\rm B-B'$ (non-stacked)            & \tisa-Ti$_{\rm S0}^{\rm A'}$ & \tisb-Ti$_{\rm S0}^{\rm B'}$ & \tisia-Ti$_{\rm S0}^{\rm A'}$ & \tisib-Ti$_{\rm S0}^{\rm B'}$ & \tisiia-Ti$_{\rm S0}^{\rm A'}$ & \tisiib-Ti$_{\rm S0}^{\rm B'}$ \\
                           & A-B (different coordination) & \tisa-\tisb  & \tisb-\tisa  & \tisia-\tisb  & \tisib-\tisa  & \tisiia-\tisb  & \tisiib-\tisa  \\ \hline
    \multirow{3}{*}{$\rm Sl-S1$} & A-A/B-B                & \tisa-\tisia  & \tisb-\tisib  & \tisia-\tisia  & \tisib-\tisib  & \tisiia-\tisia  & \tisiib-\tisib  \\
                           & A-A'/B-B'            & \tisa-Ti$_{\rm S1}^{\rm A'}$ & \tisb-Ti$_{\rm S1}^{\rm B'}$ & \tisia-Ti$_{\rm S1}^{\rm A'}$ & \tisib-Ti$_{\rm S1}^{\rm B'}$ & \tisiia-Ti$_{\rm S1}^{\rm A'}$ & \tisiib-Ti$_{\rm S1}^{\rm B'}$ \\
                           & A-B & \tisa-\tisib  & \tisb-\tisia  & \tisia-\tisib  & \tisib-\tisia  & \tisiia-\tisib  & \tisiib-\tisia  \\ \hline
    \multirow{3}{*}{$\rm Sl-S2$} & A-A/B-B                & \tisa-Ti$_{\rm S2}^{\rm A}$  & \tisb-\tisiib  & \tisia-\tisiia  & \tisib-\tisiib  & \tisiia-\tisiia  & \tisiib-\tisiib  \\
                           & A-A'/B-B'            & \tisa-Ti$_{\rm S2}^{\rm A'}$ & \tisb-Ti$_{\rm S2}^{\rm B'}$ & \tisia-Ti$_{\rm S2}^{\rm A'}$ & \tisib-Ti$_{\rm S2}^{\rm B'}$ & \tisiia-Ti$_{\rm S2}^{\rm A'}$ & \tisiib-Ti$_{\rm S2}^{\rm B'}$ \\
                           & A-B & \tisa-\tisiib  & \tisb-\tisiia  & \tisia-\tisiib  & \tisib-\tisiia  & \tisiia-\tisiib  & \tisiib-\tisiia  \\ \hline
    $\rm Sl-S0$                 & $\rm A/B-$\vo                  & \tisa-\vo           & \tisb-\vo           & \tisia-\vo           & \tisib-\vo           & \tisiia-\vo           & \tisiib-\vo          
    \end{tabular}

    \caption{Labeling scheme for the descriptors adopted in the ML model for rutile \TO(110). Each column reports the interaction categories for a polaron localized at a specific trapping site. Rows show the trapping sites of the interacting polaron or defect (\vo), including a differentiation between stacked, non-stacked, and differently coordinated (\textit{e.g.}, $\rm A$-$\rm A$, $\rm A$-$\rm A'$, $\rm A$-$\rm B$, respectively) interacting sites.}
    \label{tab:1}
\end{table}

\begin{table}[h!]
    \centering
    \small
    \setlength\tabcolsep{2pt}
    \begin{tabular}{c|l|lll}
    Layers & \multicolumn{1}{c|}{Interaction type} & \multicolumn{3}{c}{Trapping site} \\ \hline
        &   Ti$-$Ti or Ti$-$Nb  & Ti$_{\rm S0}$  & Ti$_{\rm S1} $   & Ti$_{\rm S2}$ \\ \hline
    \multirow{3}{*}{$\rm Sl-S0$} &  Ti-Ti&  Ti$_{\rm S0}$-Ti$_{\rm S0}$ & Ti$_{\rm S1}$-Ti$_{\rm S0}$  & Ti$_{\rm S2}$-Ti$_{\rm S0}$  \\
    & Ti-Nb  & Ti$_{\rm S0}$-Nb$_{\rm S0}$ & Ti$_{\rm S1}$-Nb$_{\rm S0}$ & Ti$_{\rm S2}$-Nb$_{\rm S0}$ \\ \hline
    \multirow{3}{*}{$\rm Sl-S1$} &  Ti-Ti&  Ti$_{\rm S0}$-Ti$_{\rm S1}$ & Ti$_{\rm S1}$-Ti$_{\rm S1}$  & Ti$_{\rm S2}$-Ti$_{\rm S1}$  \\
    & Ti-Nb  & Ti$_{\rm S0}$-Nb$_{\rm S1}$ & Ti$_{\rm S1}$-Nb$_{\rm S1}$ & Ti$_{\rm S2}$-Nb$_{\rm S1}$ \\ \hline
    $\rm Sl-S2$ &  Ti-Ti&  Ti$_{\rm S0}$-Ti$_{\rm S2}$ & Ti$_{\rm S1}$-Ti$_{\rm S2}$  & Ti$_{\rm S2}$-Ti$_{\rm S2}$  \\ 
    \end{tabular}
    \caption{Labeling scheme for the descriptors adopted in the ML model for \STO(001). Each column reports the interaction categories for a polaron localized at a specific trapping site. Rows show the trapping sites of the interacting polaron or defect (Nb).}
    \label{tab:2}
\end{table}
\newpage
    \subsection{Workflow}
        \begin{figure}[h!]
            \centering
            \includegraphics[width=1.0\textwidth]{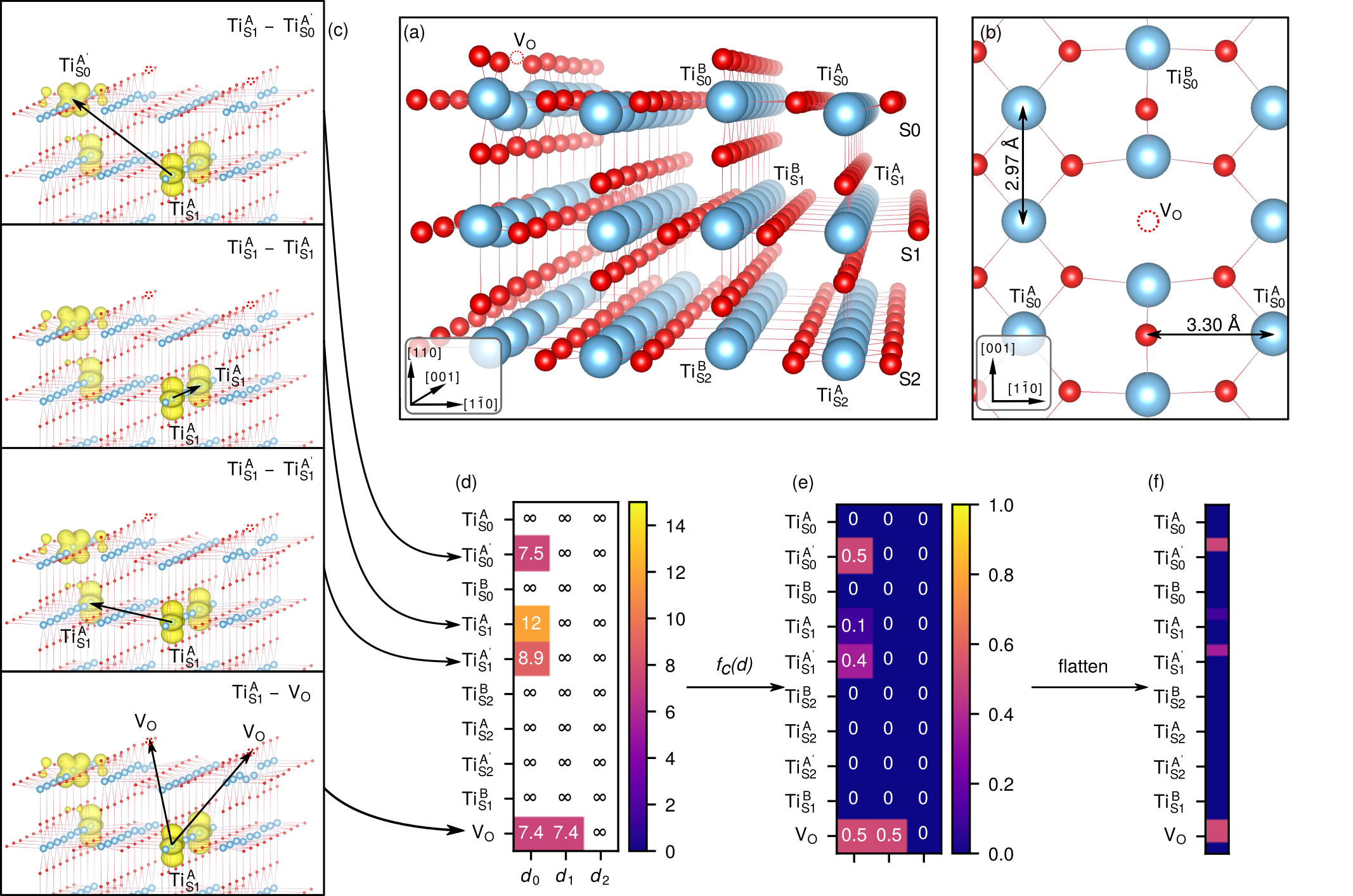}
            \caption{(a-b) Side and top view of the rutile \TO(110) structure with labeling of differently coordinated sites, distances and layers. Titanium sites are shown in blue and oxygen atoms in red. A dotted red circle shows the location of a \vo. 
            (c-f) The process of calculating the descriptor vector of a \tisia-polaron in a configuration with four polarons
            and two \vo\ (\cvo =11.1\%).
            (c) Isosurfaces of the polaron charge (yellow). Each subpanel shows a different interaction category (see labels upper right) where the target distances are indicated by an arrow.
            (d) $10\times3$ array indicating the interacting polaron-polaron and polaron-\vo\ pairs for a \tisia\ polaron according to the interactions shown in panel (c). For each interaction the polaron-polaron and polaron-\vo\  distance $d$ between the target and paired polaron/\vo\ is indicated (in \AA, also displayed with a color gradient as defined in the lateral color bar.) 
            For each interaction category only the three shortest distances (labeled $d_1$, $d_2$, and $d_3$) within a cutoff radius $R_c=15$~\AA\ are considered.  
            If too few distances are available (generally the case for low concentrations, \ie\ few polarons), the distances are padded with a value greater than $R_c$ (here $\infty$).
            (e) Illustration of the action of the rescaling function $f_c(d)$ (Equation 1 in the main text) applied to the array of distances. (f) Finally, the array is flattened into the descriptor vector $D$ of the considered polaron, used as input in our machine learning model, as described in the text.
            }
            \label{fig:structure}
        \end{figure}
        
    Figure~\ref{fig:structure} shows the workflow for determining a polaron descriptor, considering a specific polaron configuration on \TO(110).
    First, we recall the structural unit of rutile \TO(110) and the corresponding labeling in Fig.~\ref{fig:structure}(a-b). 
    Fig.~\ref{fig:structure}(c) shows a polaron configuration and all active interaction categories for the central polaron located at a \tisia-site. In Fig.~\ref{fig:structure}(d), the corresponding 2-body-interactions are determined and structured according to the interaction categories. After rescaling (Fig.~\ref{fig:structure}(e)) the array is flattened (Fig.~\ref{fig:structure}(f)) into a vector for further computations.

\newpage

\clearpage

\section{Model optimization and evaluation}
    In the following we show the convergence of the models in the training procedure. The hyperparameter optimization results are shown for the case of rutile \TO(110). Finally, we show scatter plots of the omitted defect concentrations cases and provide quantitative information on errors on the target quantity $\bar{E}_{\rm pol}$.
    
    \subsection{Training}
    We recorded the adapted loss (see Equation~3 in the main text) at an interval of 100 stochastic optimization epochs for the training and validation phases, using a randomized split of the databases.
    \begin{figure}[h!]
        \centering
        \includegraphics[width=\textwidth]{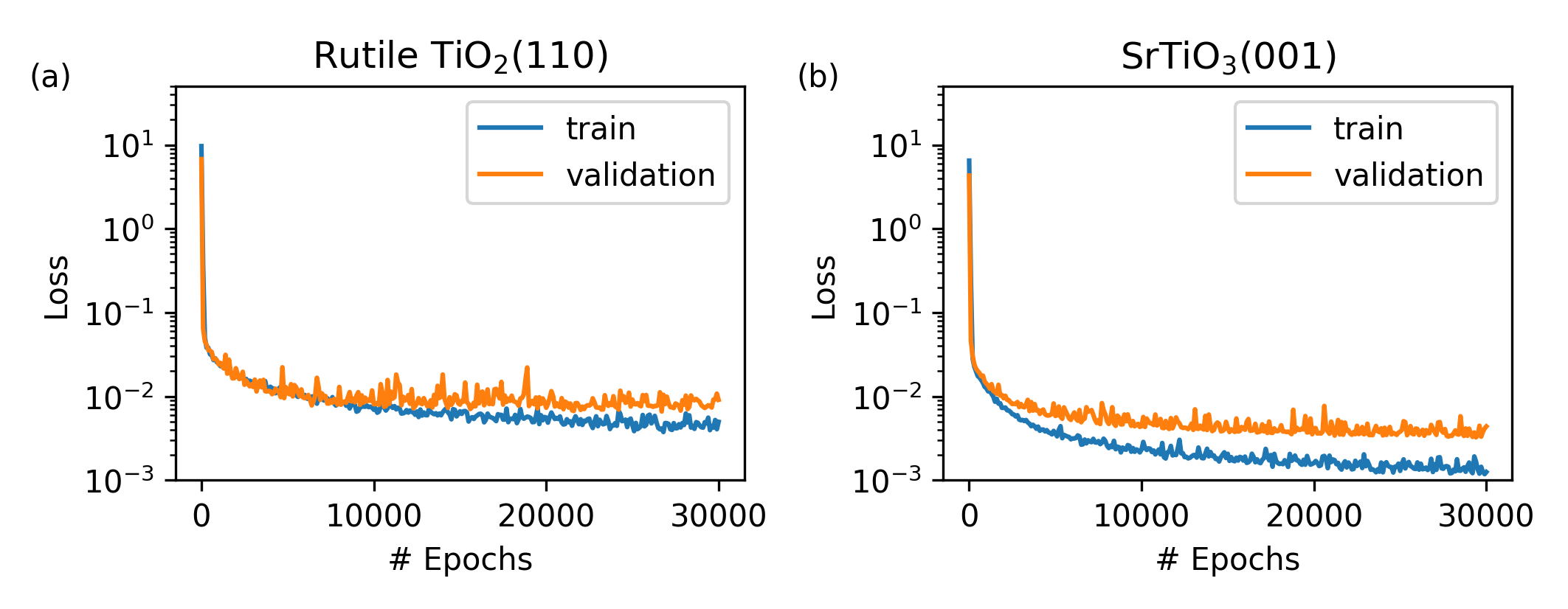}
        \caption{The adapted loss function is shown for rutile \TO\ (a) and \STO\ (b) in dependence of the number of trained epochs in the stochastic parameter optimization.}
        \label{fig:learning}
    \end{figure}

\clearpage

\subsection{Hyperparameter optimization}
\begin{figure*}[h!]
        \centering
        \begin{subfigure}[b]{0.475\textwidth}
            \centering
            \includegraphics[width=\textwidth]{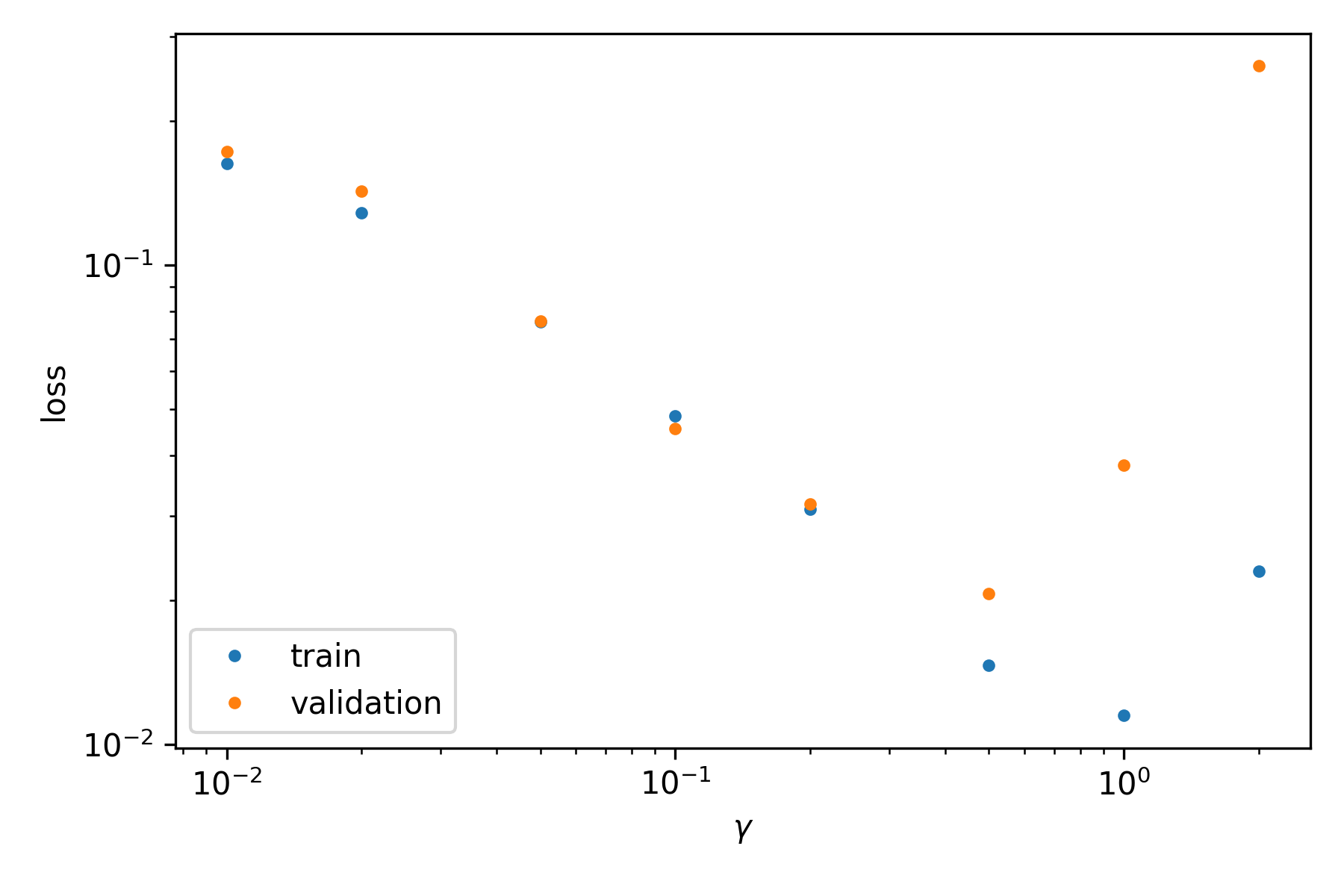}
            \caption[]%
            {{\small Loss of train and validation data obtained after 5000 epochs, usging $R_c=15$~\AA\ and various values $\gamma$ of the Laplacian kernel. }}    
            \label{fig:mean and std of net14}
        \end{subfigure}
        \hfill
        \begin{subfigure}[b]{0.475\textwidth}  
            \centering 
            \includegraphics[width=\textwidth]{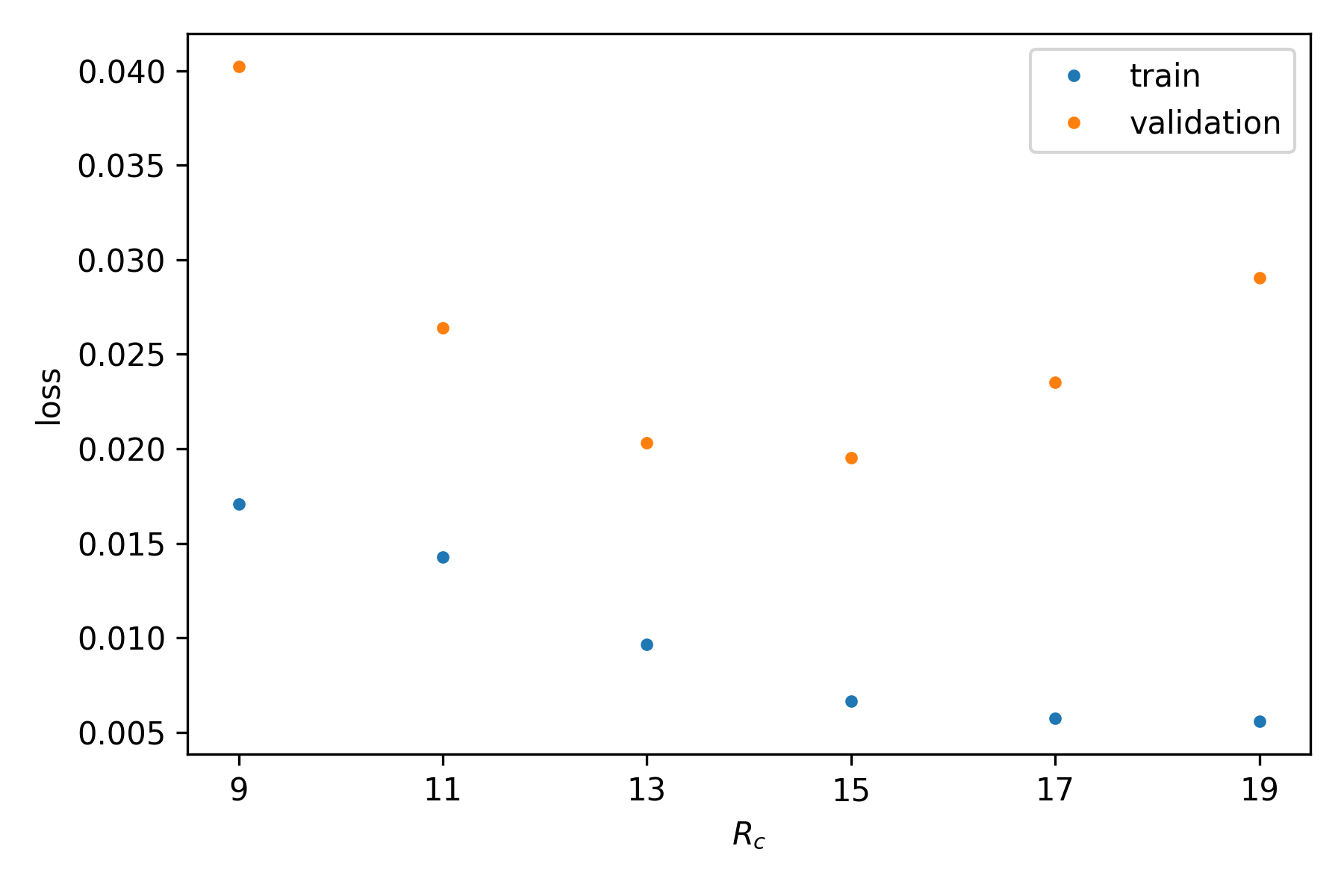}
            \caption[]%
            {{\small Loss of train and validation data obtained after 10000 epochs, using $\gamma=0.5$ and various $R_c$.}}    
            \label{fig:mean and std of net24}
        \end{subfigure}
        \vskip\baselineskip
        \begin{subfigure}[b]{0.475\textwidth}   
            \centering 
            \includegraphics[width=\textwidth]{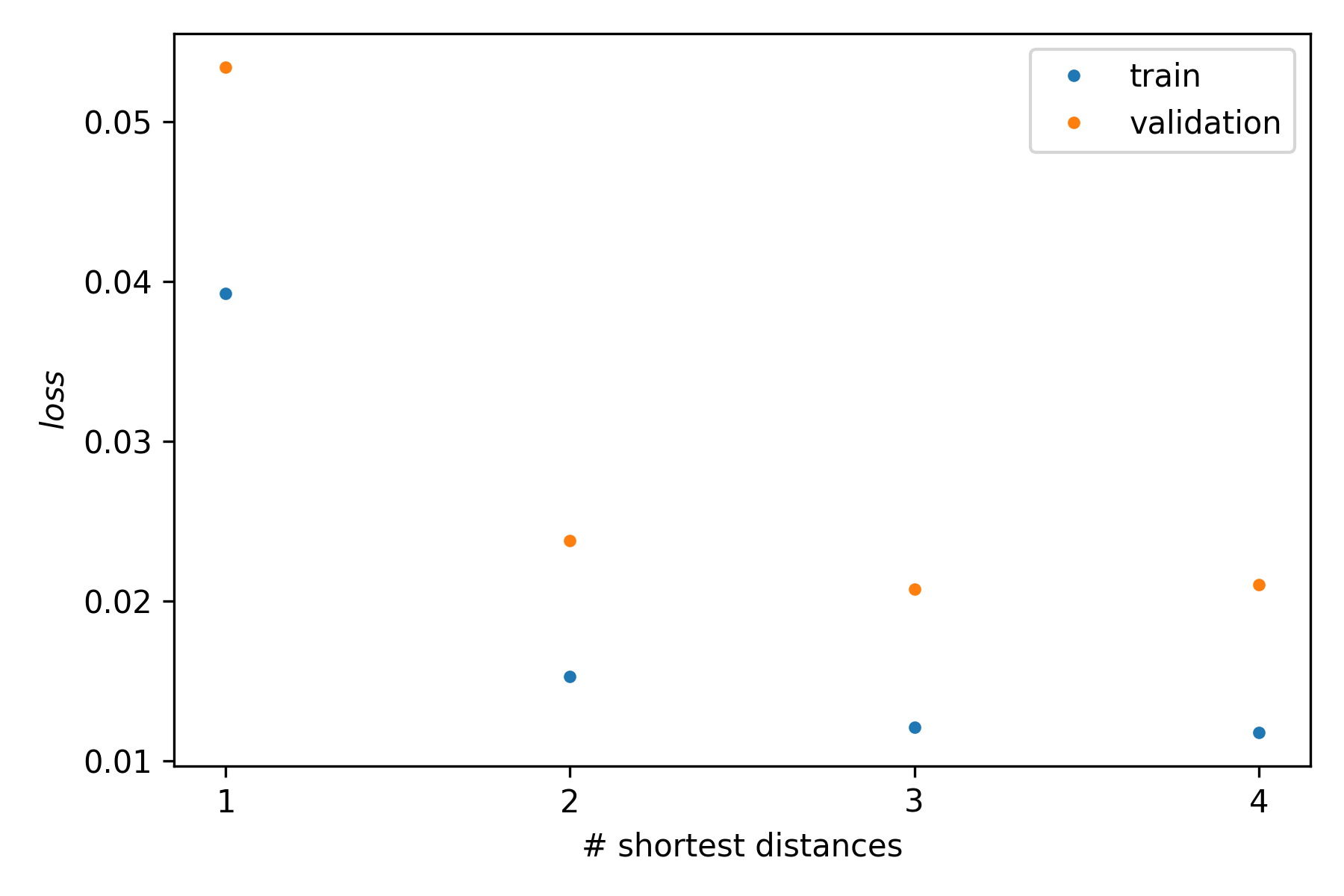}
            \caption[]%
            {{\small  Loss of train and validation data obtained after 5000 epochs for $\gamma=0.5$, by varying the number of shortest distances included in each interaction category. While different interaction categories could be tuned separately to further optimize the descriptor, here we limited the optimization to an equal number of features per interaction category.}}    
            \label{fig:mean and std of net34}
        \end{subfigure}
        \hfill
        \caption[]
        {\small Examples of the hyperparameter optimization based of results from oxygen defective rutile \TO(110).} 
        \label{fig:hyperpara}
    \end{figure*}

     We determined optimal hyperparameters of the model (kernel parameter $\gamma$, cutoff-radius $R_c$ and number of elements per interaction category) by optimizing the validation accuracy. The results of this procedure are displayed for rutile \TO(110) in Figure \ref{fig:hyperpara}.
     
\clearpage

\subsection{Omitted defect concentrations}
\begin{figure}[h!]
    \centering
    \includegraphics[width=0.65\textwidth]{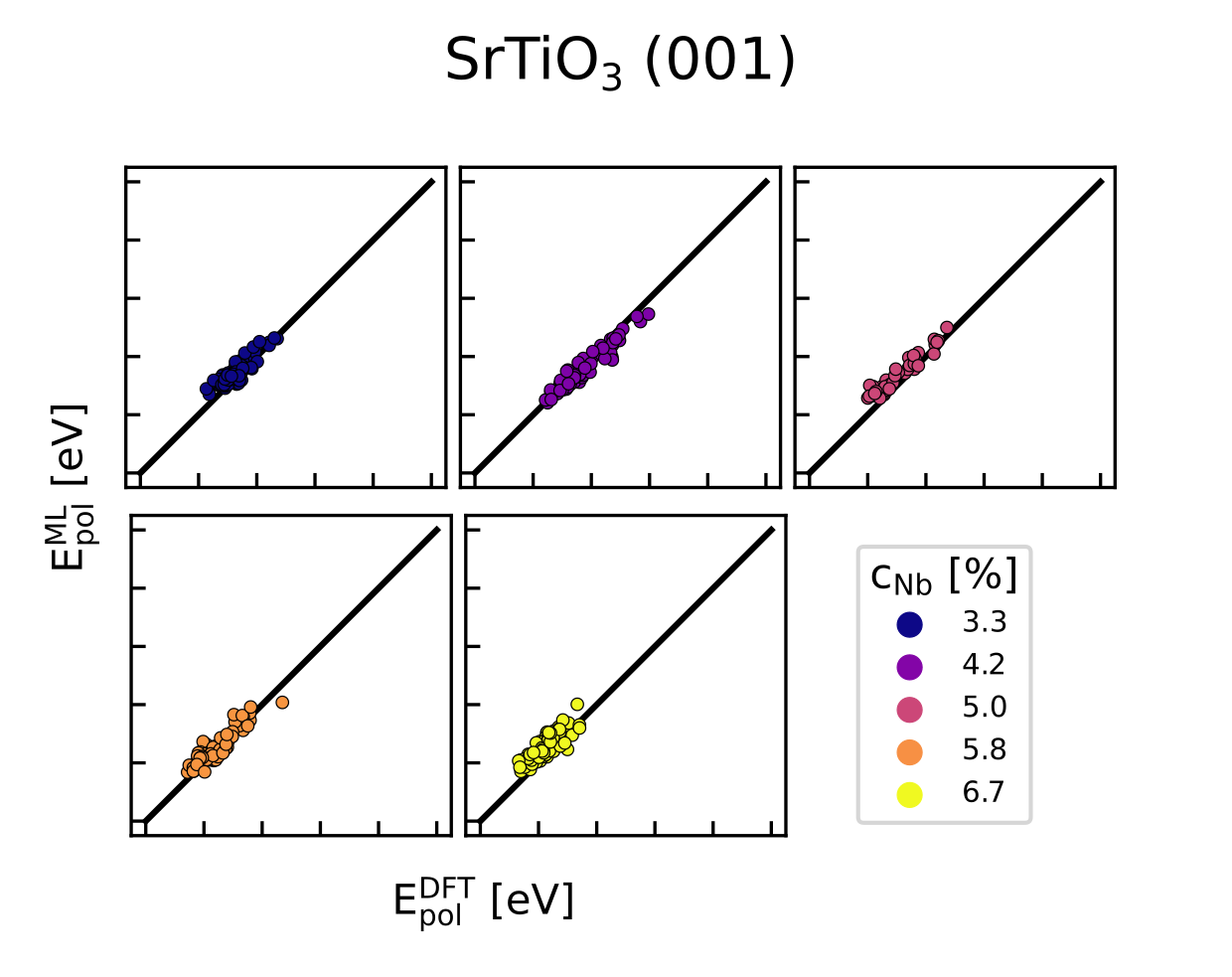}
    \includegraphics[width=0.65\textwidth]{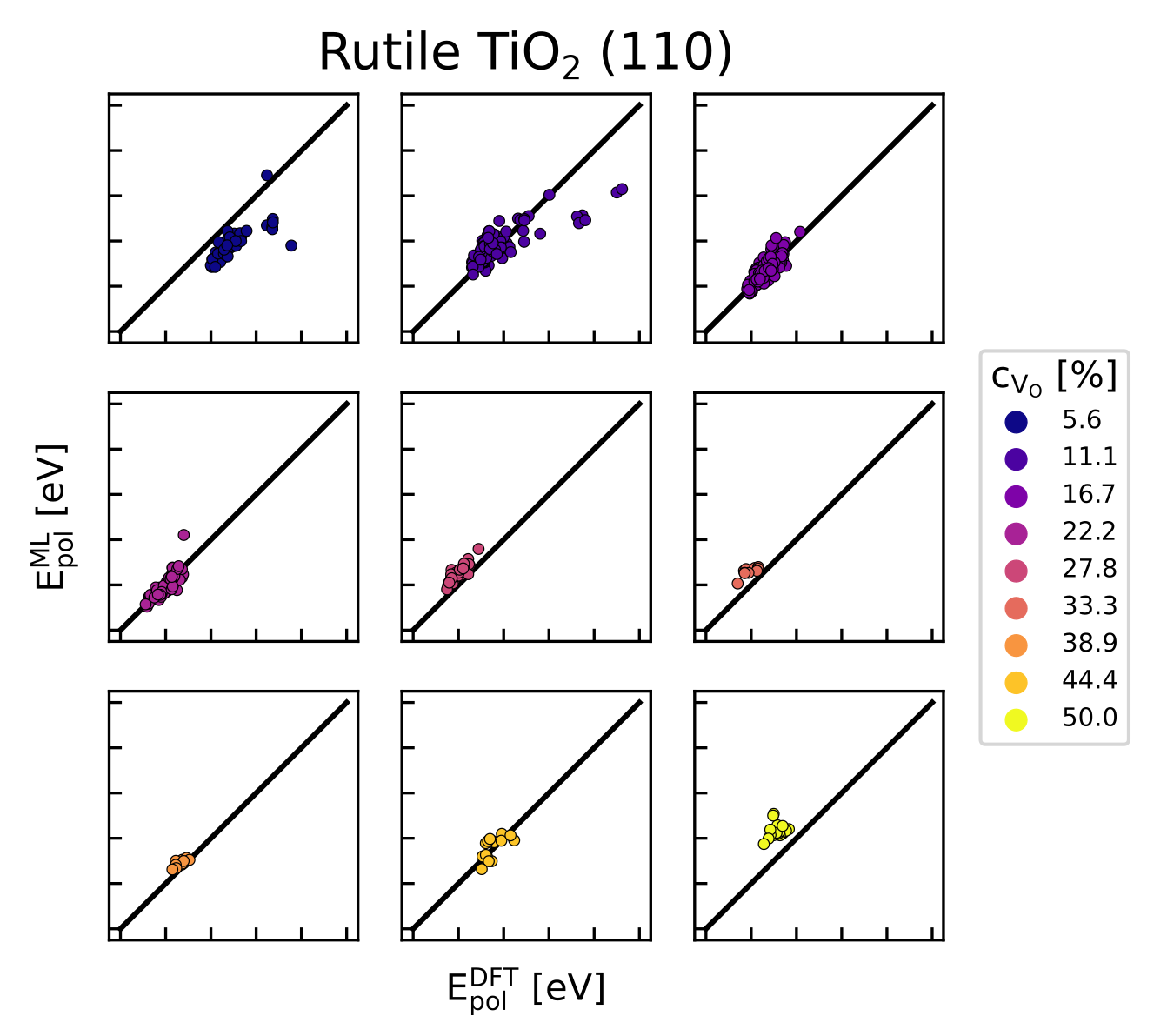}
    \caption{The results of the validation of interpolation to omitted defect concentrations are shown for \STO\ and \TO. Each panel displays only the results of the testing data from the omitted defect concentration, while the model was trained on data of all remaining defect concentrations. The color of the points encodes the defect concentration of the corresponding data.}
    \label{fig:omitted}
\end{figure}

    Considering that the goal of this work is the efficient exploration of the polaron configurational space, the ML-model must be robust and guarantee accurate predictions for samples not visited in the training phase.
    Usually, the application of ML-models for extrapolation problems results in unreliable predictions, and careful tests are required.
    To this aim, we adopted a validation scheme, the ``omitted defect concentrations'' presented in the main text.
     Figure \ref{fig:omitted} shows the results at every defect concentration for both the \STO\ and \TO\ systems.
     Furthermore, in Table \ref{tab:loss} and \ref{tab:loss_nb}, we report the mean squared error of the target quantity $\bar{E}_{\rm pol}$ as obtained in the training (including all concentrations in the database except the omitted one) and the test phase (omitted concentration only).
     The results show the good level of accuracy of the proposed ML model in addressing novel polaron configurations. The error obtained for the omitted concentration is indeed slightly larger than the value obtained for the standard test on randomly split training-vs-test databases (see last line in the Tables).

    \begin{table}[h!]
        \centering
        \begin{tabular}{c | c | c }
        \cvo & Train & Test \\
        \hline \hline
        5.5\%  & $1.11\cdot10^{-4}$  & $2.61\cdot10^{-3}$    \\
        11.1\% & $1.21\cdot10^{-4}$  & $1.72\cdot10^{-3}$    \\
        16.7\% & $1.64\cdot10^{-4}$  & $4.85\cdot10^{-4}$   \\
        22.2\% & $2.06\cdot10^{-4}$  & $2.65\cdot10^{-4}$   \\
        27.8\% & $2.43\cdot10^{-4}$  & $5.86\cdot10^{-4}$    \\
        33.3\% & $1.71\cdot10^{-4}$  & $1.45\cdot10^{-3}$    \\
        38.9\% & $1.97\cdot10^{-4}$  & $5.32\cdot10^{-4}$    \\
        44.4\% & $2.07\cdot10^{-4}$  & $5.27\cdot10^{-4}$    \\
        50\%   & $1.74\cdot10^{-4}$  & $3.04\cdot10^{-3}$               \\
        \hline \hline
        Randomized & $1.16\cdot10^{-4}$ &  $1.31\cdot10^{-4}$  \\
        \end{tabular}
        \caption{The mean squared error of the mean polaronic energy for different test cases in the MD-dataset, reported for the training and test sets. The training was performed on configurations from eight defect concentrations, and the testing on the remaining one (labeled in column \cvo). 
        The last line shows results from a randomized split of the data from all defect concentrations.
        }
        \label{tab:loss}
    \end{table}
    
    \begin{table}[h!]
        \centering
        \begin{tabular}{c | c | c }
        $c_{\rm Nb}$ & Train & Test \\
        \hline \hline
        3.3\%  & $3.01\cdot10^{-5}$  & $2.19\cdot10^{-4}$    \\
        4.2\% & $4.26\cdot10^{-5}$  & $1.64\cdot10^{-4}$    \\
        5.0\% & $3.25\cdot10^{-5}$  & $3.79\cdot10^{-4}$   \\
        5.8\% & $3.64\cdot10^{-5}$  & $1.75\cdot10^{-4}$   \\
        6.7\% & $2.92\cdot10^{-5}$  & $4.07\cdot10^{-4}$    \\
        \hline \hline
        Randomized & $5.75\cdot10^{-5}$ &  $9.35\cdot10^{-5}$  \\
        \end{tabular}
        \caption{The mean squared error of the mean polaronic energy for different test cases in the randomized dataset, reported for the training and test sets. The training was performed on configurations from 4 defect concentrations, and the testing on the remaining one (labeled in column $c_{\rm Nb}$). 
        The last line shows results from a randomized split of the data from all defect concentrations.
        }
        \label{tab:loss_nb}
    \end{table}

\clearpage

\section{Exhaustive Search}
\begin{figure}
        \centering
        \includegraphics[width=0.85\textwidth]{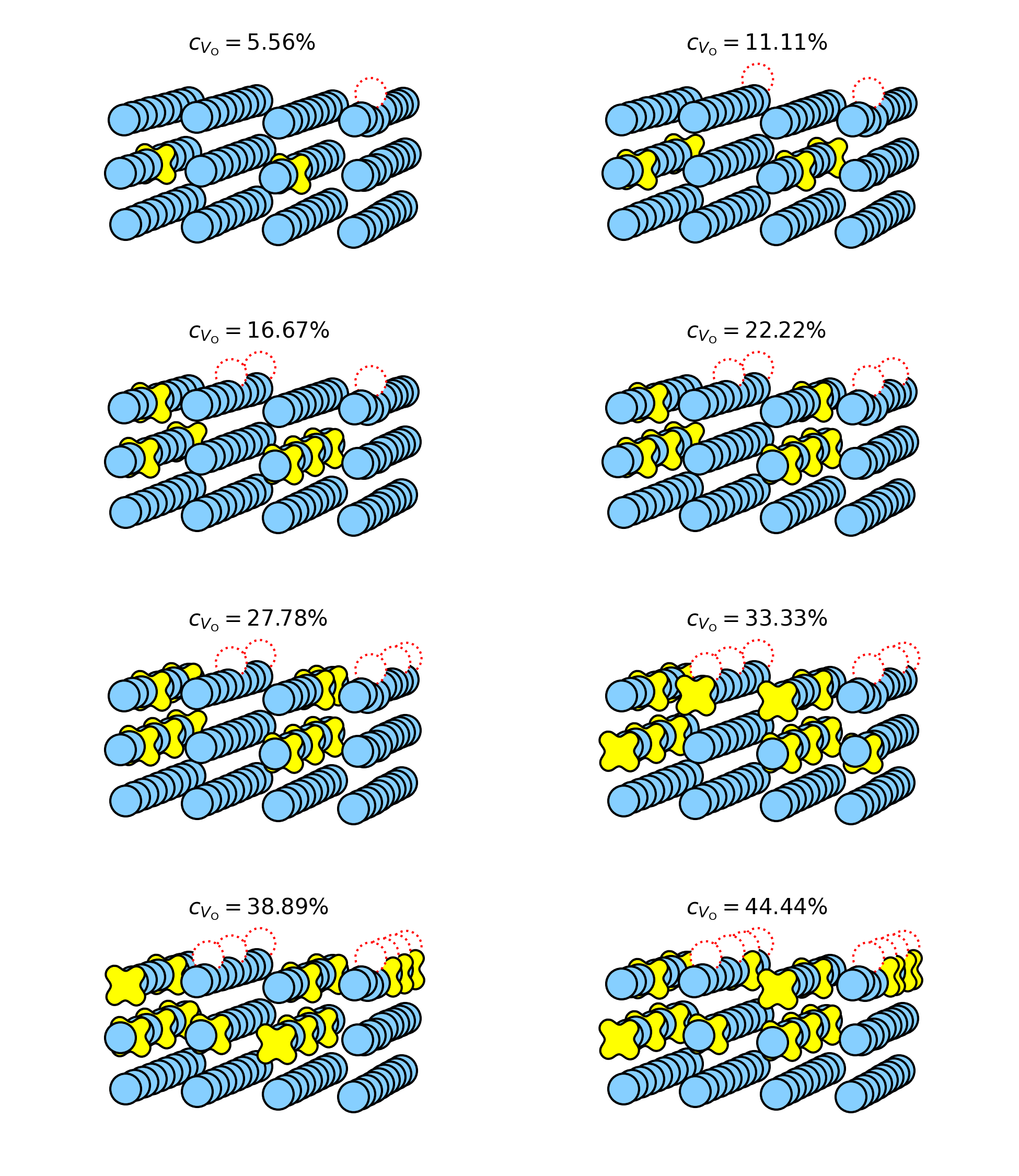}
        \caption{Most stable polaron configurations predicted by ML and confirmed at the DFT-level at each \cvo\ for rutile \TO(110). Titanium-sites are shown in blue, polarons are indicated in yellow by a schematic representation of the charge density, and oxygen vacancies are shown in dotted red. For clarity oxygen atoms are not shown and only the three surface layers are displayed.}
        \label{fig:most_fav_to}
    \end{figure}

\begin{figure}
        \centering
        \includegraphics[width=0.85\textwidth]{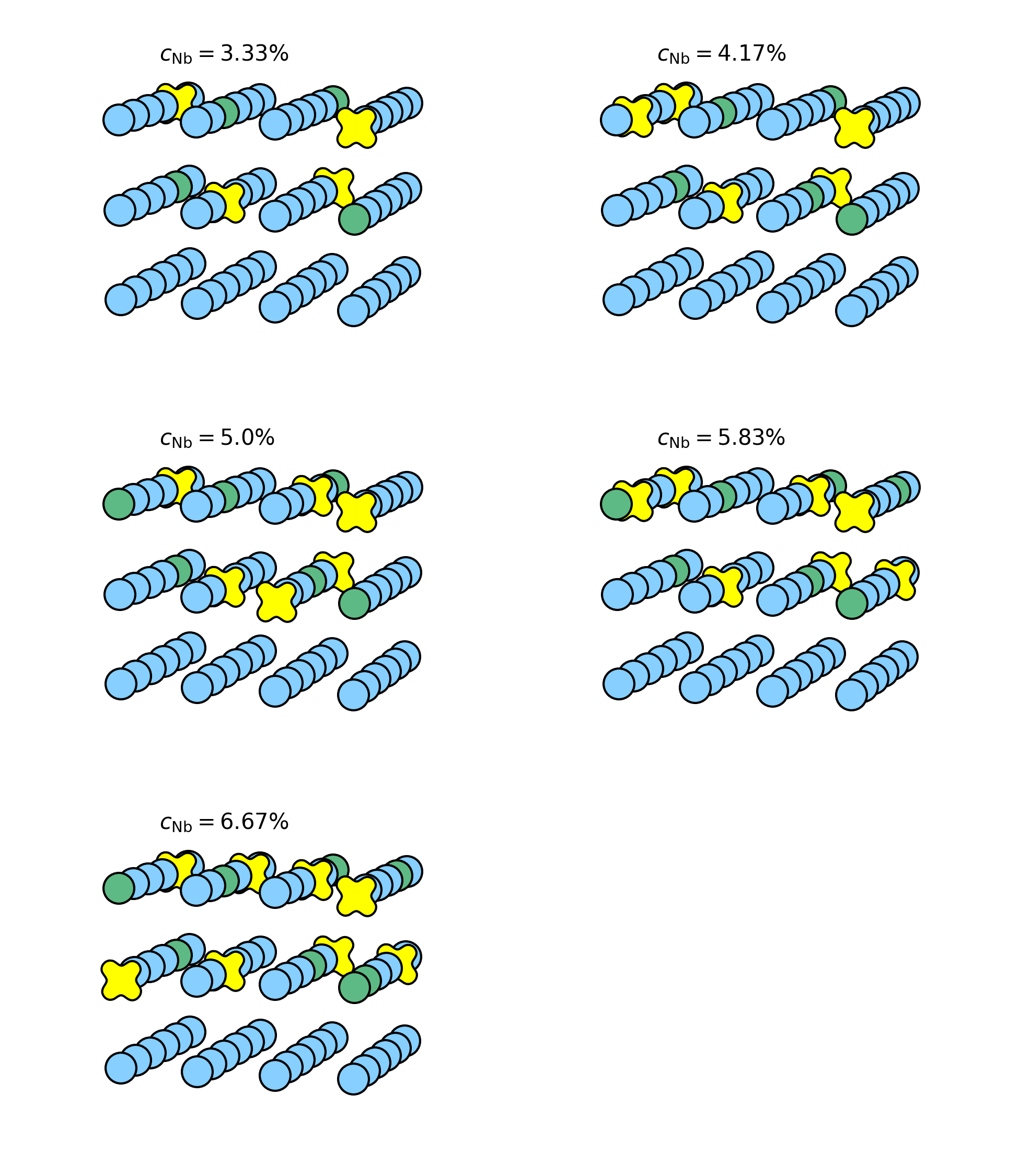}
        \caption{Most stable polaron configurations predicted by ML and confirmed at the DFT-level at each \cnb\ for \STO(001). Titanium-sites are shown in blue, polarons are indicated in yellow by a schematic representation of the charge density, and Nb-dopants are shown in green. For clarity oxygen and strontium atoms are not shown and only three surface layers are displayed.}
        \label{fig:most_fav_sto}
    \end{figure}

\begin{figure}
    \centering
    \includegraphics[width=0.85\textwidth]{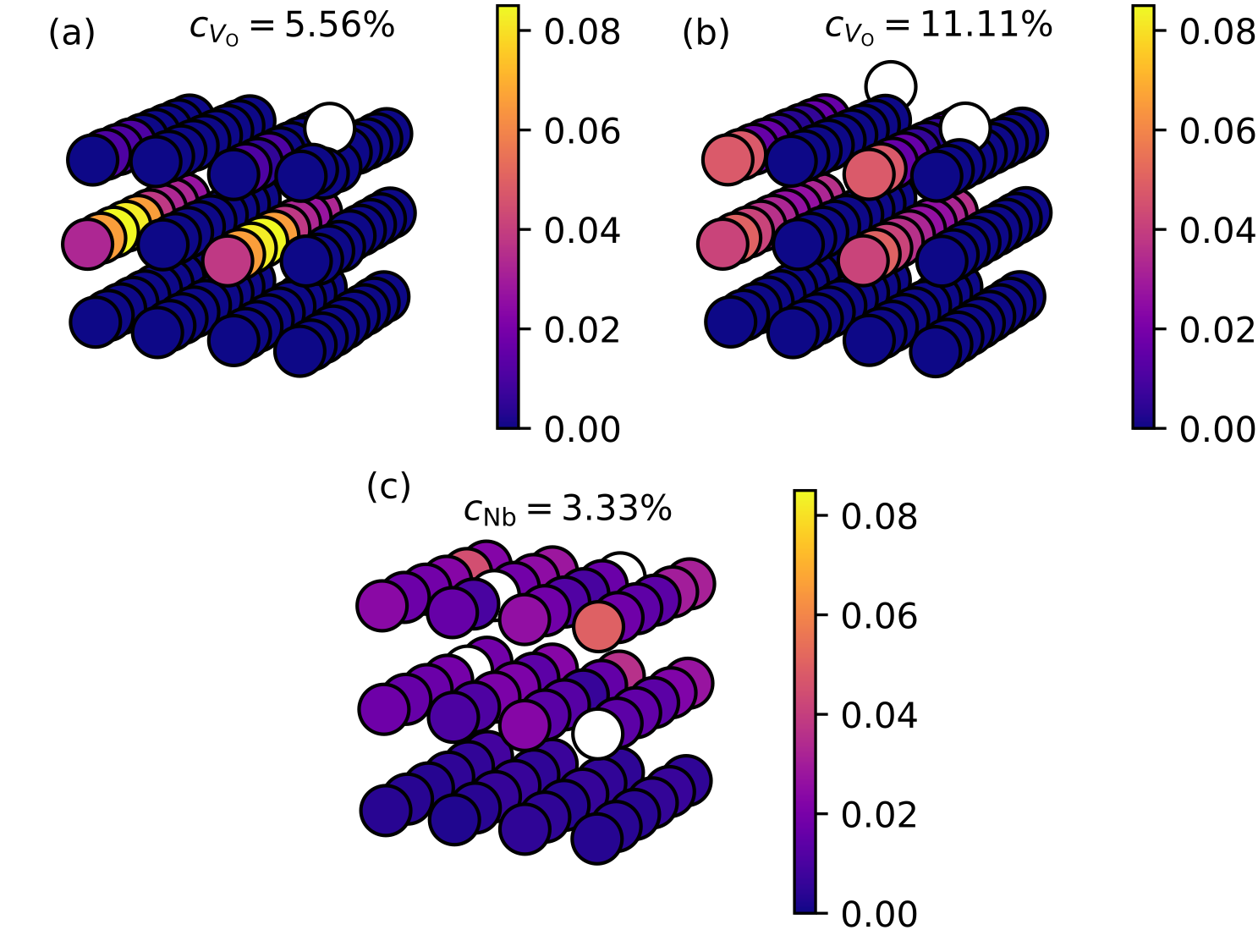}
    \caption{The relative site occupation frequency within the most favorable, ML-searched configurations are displayed for the (a,b) rutile \TO(110) and (c) perovskite \STO(001) surface, respectively. The color encoding and the corresponding lateral colorbars show the percentage of favorable configurations containing a polaron at each Ti-site in the supercell. The positions of defects (\vo\ and Nb-dopants, respectively) are indicated by a white circle.}
    \label{fig:dist_to_sto}
\end{figure}

Figure~\ref{fig:most_fav_to} and Figure~\ref{fig:most_fav_sto} collect the most-stable polaronic configuration as found by the machine-learning exhaustive search for every defect concentration in \TO\ and \STO, respectively.
These ground-state configurations were individuated by the ML algorithm among a pool of $\sim 4$ and $\sim 2$~million polaron patterns, respectively.
We note that searches based purely on DFT calculations are limited to smaller datasets due to the higher computational costs:
The characterization of just $\sim 500$ distinct polaronic configurations in \TO\ (\ie\ calculations including MD simulations plus ionic relaxations for the calculation of \EPOLDFT) required us $\sim 2 \cdot 10^6$ core hours on our high-performance computing facility (HPC);
similarly, the $\sim 400$ configurations analyzed for \STO\ via the random sampling approach were calculated in $\sim 0.3 \cdot 10^6$ HPC-core hours.
The exhaustive ML-search on a configuration space of millions of polaron pattern requires instead a handful of hours (11 and 2~hours for \TO\ and \STO, respectively) on a single core of standard personal computers.
The bottleneck of the ML-based strategy is the training data generation, which relies on DFT databases;
however, our results for \STO\ show that the ML model can be efficiently trained by performing a random sampling of the configuration space, relying on few hundreds of polaron energies calculated via DFT.
The machine learning workflow proposed in this work is indeed able to overcome the practical limitations of approaches based purely on DFT and on physical intuition.
We conclude with a statistical analysis on the polaron patters explored during the exhaustive search for \TO\ and \STO\ at low defect concentrations, as shown in Figure~\ref{fig:dist_to_sto}.
By considering only favorable configurations (\ie\ configurations with \EPOLML\ at least 80\% of the ground state energy), the frequency distribution of polarons in \TO\ shows a clear tendency towards charge trapping in \tisia\ sites close to the \vo\ at the lowest low defect concentration (\ref{fig:dist_to_sto} (a)), correctly capturing the underlying symmetries of the supercell.
With increasing defect concentration (see \ref{fig:dist_to_sto} (b)), polarons tend to distribute more evenly across all \tisia\ sites compared to the lower concentration, and occupation of \tisa\ sites between the two \vo\ becomes more favorable.
In the case of an asymmetric defect pattern, as for \STO, the distribution of favorable sites gets more complicated (see \ref{fig:dist_to_sto} (c)).
Nevertheless, The ML-aided search identifies sites directly above or below the Nb-dopants in the first two surface layers as the preferred polaron localization sites.

\clearpage
\bibliographystyle{plain}
\bibliography{references}